\documentclass[english]{article}
\usepackage[T1]{fontenc}
\usepackage[latin9]{inputenc}
\usepackage{geometry}
\usepackage{float}
\usepackage{amsmath}
\usepackage{esint}

\makeatletter

\providecommand{\tabularnewline}{\\}

\@ifundefined{date}{}{\date{}}
\numberwithin{equation}{section}

\makeatother

\usepackage{babel}
\begin{document}

\title{Three-body force for baryons from the D0-D4/D8 matrix model}

\maketitle
\begin{center}
Si-wen Li\footnote{Email: cloudk@mail.ustc.edu.cn} and Tuo Jia\footnote{Email: jt2011@mail.ustc.edu.cn}
\par\end{center}

\begin{center}
\emph{Department of Modern Physics, }\\
\emph{ University of Science and Technology of China, }\\
\emph{ Hefei 230026, Anhui, China}
\par\end{center}

\vspace{10mm}

\begin{abstract}
This is an extensive work to our previous paper \cite{key-08} studied
on the D0-D4/D8 holographic system. We compute the three-body force
for baryons with the D0-D4/D8 matrix model derived in \cite{key-08}
with considering the non-zero QCD vacuum. We obtain the three-body
force at short distances but modified by the appearance of the smeared
D0-branes i.e. considering the effects from the non-trivial QCD vacuum.
We firstly test our matrix model in the case of 't Hooft instanton
and then in two more realistic case: (1) three-neutrons with averaged
spins and (2) proton-proton-neutron (or proton-neutron-proton). The
three-body potential vanishes in the former case while in two latter
cases it is positive i.e. repulsive and makes sense only if the constraint
for stable baryonic state is satisfied. We require all the baryons
in our computation aligned on a line. These may indicate that the
cases in dense states of neutrons such as in neutron stars, Helium-3
or Tritium nucleus all with the non-trivial QCD vacuum.
\end{abstract}
\newpage{}

\section{Introduction}

In nuclear physics, there is one of the fundamental ingredients which
is how to describe the interaction among nucleons. QCD as an underlying
theory of strong interactions with asymptotic freedom, makes it possible
to form the confinement such as bound states of nucleons. People have
to study on the behavior of nucleons to describe the nuclear force.
As it is known, the nuclear force can not only be explained by two-body
force, but also three-body force which plays the important role as
well. For example, the three-body nuclear force is vital in studying
on the excitation spectra of light nuclei or high-density baryon matters
such as supernovae or neutron stars. However the properties of three-body
nuclear force are still to be revealed although it has been developed
so many years. Since QCD at nonzero baryon density is strongly coupled
which thus is notoriously difficult to solve, consequently it becomes
the main obstacle for revealing the aspects of nuclear force.

On the other hand, some classical or semi-classical gauge field configurations
may also be important in QCD or nuclear physics, for example, some
topologically nontrivial solutions such as instantons, monopoles and
so on. In \cite{key-01,key-02,key-03,key-04}, the self-dual field
strength has been studied and proposed to be a mechanism for the confinement
\cite{key-05}. Therefore, the states with non-zero QCD vacuum $\theta$
angle (or equivalently non-zero $\mathrm{\theta Tr}\left(F_{\mu\nu}\tilde{F}^{\mu\nu}\right)$,
here $\tilde{F}^{\mu\nu}$ is the dual field of the gauge field strength
$F_{\mu\nu}$.) could exist and affect the the mass of meson with
the interaction among baryons, and this $\theta$-dependent term may
also lead to some other observable effects such as in the glueball
condensation or the chiral magnetic effect (CME) \cite{key-06,key-07}.
Thus in this paper, we would like to study on the three-body nuclear
force with non-zero QCD vacuum $\theta$ angle by using the D0-D4/D8
holographic matrix model proposed in \cite{key-08} for baryons. We
compute the three-body force at short distances for baryons in the
large $N_{c}$ holographic QCD explicitly while the two-body case
has already been studied in \cite{key-08}. 

By holography, in order to describe the states with non-zero $\mathrm{Tr}\left(F_{\mu\nu}\tilde{F}^{\mu\nu}\right)$
in the dual field theory, it corresponds to adding smeared D0-branes
to the compacted D4-brane background \cite{key-09,key-10}. And with
introducing the flavor $\mathrm{D}8/\overline{\mathrm{D}8}$-branes,
the meson spectrum has been studied in \cite{key-09} while the baryon
spectrum was studied in \cite{key-11} with the approach of Yang-Mills
instanton. As a comparison with \cite{key-11} and study from \cite{key-12},
we use the gauge/string duality (or AdS/CFT correspondence, see \cite{key-13,key-14,key-15}
for a review.) to derive our matrix model from the Sakai-Sugimoto
model \cite{key-16,key-17} in the D0-D4 background \cite{key-09}
(i.e. D0-D4/D8 system), in the large-$N_{c}$ limit at a large 't
Hooft coupling $\lambda$. That is to say, our matrix model is also
a low-energy effective theory on the baryon vertex, which is the D4'-branes\footnote{In order to distinguish the D4-branes which are responsible for the
background geometry, we use D4'-brane to denote the baryon vertex
since in D0-D4/D8 system the baryon vertex is also a D4-brane.} wrapped on $S^{4}$ in D0-D4/D8 system \cite{key-09,key-18}, in
the gravity side of gauge/string duality. 

The matrix model could describe $k$-body baryons with arbitrary $k$,
since the rank of the matrix is determined by the value of $k$, where
$k$ is the baryon number. In our matrix model, the positions of $k$
baryons are represented by the diagonal elements of the matrices after
integrating out the off-diagonal elements. Furthermore, the classical
values of a pair of the complex $k\times N_{f}$ rectangular matrices
are related to the sizes of baryons and they describe the dynamics
of the strings connecting the flavor $\mathrm{D}8/\overline{\mathrm{D}8}$-branes
and the baryon vertices. With all of these, it comes to the well-known
Atiyah-Drinfeld-Hitchin-Manin (ADHM) matrix model of instantons.

In our previous work \cite{key-08}, we also studied on the baryon
spectrum ($k=1$) and two-body force at short distances for baryons
($k=2$), except the derivation of our matrix model. For the case
of $k=1$, we find a constraint for stable baryonic state in D0-D4/D8
system which is exactly the same as the conclusion in \cite{key-11}
with the approach of Yang-Mills instanton, but quite different from
the original Sakai-Sugimoto model. And our baryon spectrum could fit
the experimental data well just by adjusting the number density of
the D0-branes. For the case of $k=2$, we have studied on the two-body
force at short distances and also found a universal repulsive core
for any baryonic state but modified by the appearance of the smeared
D0-branes. It turns out that the two-body force at short distances
could become attractive which describes an unstable two-body system
if the constraint for stable baryonic state is overcome.

Instead of phenomenological models, our matrix model is for multi-bayon
systems with non-zero QCD vacuum $\theta$ angle and based on the
gauge/string duality with the underlying string theory, so it is natural
and interesting to extend the analysis to derive the three-body force
in D0-D4/D8 system with our matrix model. So in this paper, we extend
our previous work in \cite{key-08} and continue the analysis to the
case of $k=3$, as a parallel computation to \cite{key-19}, to study
the three-body force at short distances with non-trivial QCD vacuum
by using our matrix model. We will focus on the two particular examples
which are three neutrons with averaged spins and proton-proton-neutron
(or proton-neutron-neutron), and require all the baryons or nucleons
aligned on a line with equal spacings for each case. The system with
averaged spins is typical for dense states of multi-baryons in QCD,
such as cores of neutron stars, while the latter one is related to
tritium nuclei or Helium-3. From our results, we find the three-body
potential is suppressed if compared to the two-body force in \cite{key-08}.
And both in the case of averaged spins and proton-proton-neutron,
the three-body potential would be totally complex if $\zeta=U_{Q_{0}}^{3}/U_{KK}^{3}>2$
where $U_{Q_{0}}^{3}$ is related to the number density of the smeared
D0-brane. This result is exactly the same as the constraint for the
stable baryonic state claimed in \cite{key-08} and \cite{key-11}
with the approach of Yang-Mills instanton.

In this paper, the organization is as follows. In section 2, we briefly
review the D0-D4/D8 matrix model and the calculations for two-body
force as shown in \cite{key-08}. In section 3, we calculate the three-body
force with the ``classical treatment'' i.e. the case with spin or
isospin aligned classically. In this case we find a vanished three-body
force which is independent on the non-zero QCD vacuum $\theta$ angle\footnote{In fact, our result depends on the parameter $\zeta$ in this holographic
model, however it has been turned out $\zeta$ is related to the parameter
$\theta$ in the topological term in QCD \cite{key-09}. }, however this result is similar as in \cite{key-19} and consistent
with the soliton approach in \cite{key-20}. Then we employ the set-up
for generic three-body forces with quantum spin/isospin, the resultant
three-body force is modified by the appearance of the smeared D0-branes
(or equivalently by considering the non-trivial QCD vacuum) and also
consistent with the constraint for stable baryonic states claimed
in \cite{key-08,key-11}. The summary and discussion are in the final
section.

\section{A brief review of D0-D4/D8 matrix model}

To calculate the three-body force for baryons by using D0-D4/D8 matrix
model is quite analogous to the computation of the two-body force
which has been performed in \cite{key-08}. In this section, we will
give a brief summary of the D0-D4/D8 matrix model and the calculations
of the two-body force for baryons with this model.

\subsection{Action of D0-D4/D8 matrix model}

We proposed a D0-D4/D8 matrix model in \cite{key-08} by using the
standard technique in string theory from the Sakai-Sugimoto model
in the D0-D4 background (i.e. D0-D4/D8 system). It is a quantum mechanical
system with $U\left(k\right)$ symmetry which takes the following
action

\begin{eqnarray}
S & = & \frac{\lambda N_{c}M_{KK}}{54\pi}\left(1+\zeta\right)^{3/2}\mathrm{Tr}\int dt\bigg[\left(D_{0}X^{M}\right)^{2}-\frac{2}{3}\left(1-\frac{1}{2}\zeta\right)M_{KK}^{2}\left(X^{4}\right)^{2}\nonumber \\
 &  & +D_{0}\bar{\omega}_{i}^{\dot{\alpha}}D_{0}\omega_{i\dot{\alpha}}-\frac{1}{6}\left(1-\frac{1}{2}\zeta\right)M_{KK}^{2}\bar{\omega}_{i}^{\dot{\alpha}}\omega_{i\dot{\alpha}}\nonumber \\
 &  & +\frac{3^{6}\pi^{2}}{4\lambda^{2}M_{KK}^{4}}\frac{1}{\left(1+\zeta\right)^{4}}\left(\vec{D}\right)^{2}+\vec{D}\cdot\vec{\tau}_{\ \dot{\beta}}^{\dot{\alpha}}\bar{X}^{\dot{\beta}\alpha}X_{\dot{\alpha}\alpha}+\vec{D}\cdot\vec{\tau}_{\ \dot{\beta}}^{\dot{\alpha}}\bar{\omega}^{\dot{\beta}\alpha}\omega_{\dot{\alpha}\alpha}\bigg]\nonumber \\
 &  & +N_{c}\mathrm{Tr}\int dtA_{0}\ .\label{eq:Matrix action}
\end{eqnarray}

\noindent It is allowed to change the baryon number $k$ by choosing
the gauge group $U\left(k\right)$ for the matrix model (\ref{eq:Matrix action})
to describe $k$-body interaction in the D0-D4/D8 system. Note that
the rank of the gauge group $U\left(k\right)$ is the number of baryons.
The parameter $\lambda=g_{YM}^{2}N_{c}$ is the 't Hooft coupling
constant and $M_{KK}$ is a unique scale while the parameter $\zeta$
is $\zeta=U_{Q_{0}}^{3}/U_{KK}^{3}$, where $U_{Q_{0}}^{3}$ is related
to the number density of smeared D0-branes. $N_{c}$ and $N_{f}$
represent the number of colors and flavors respectively. We have obtained
the baryon spectrum for the case of $k=1$ and fitted the experimental
data by adjusting the parameter $\zeta$, computed the two-body force
(i.e. $k=2$) for baryons at short distances with the matrix model
(\ref{eq:Matrix action}) in \cite{key-08}. To clarify the symmetry
in the matrix model (\ref{eq:Matrix action}) , we summarize the field
content in the following table.

\noindent 
\begin{table}[H]
\noindent \begin{centering}
\begin{tabular}{|c|c|c|c|c|}
\hline 
Fields & index & $U\left(k\right)$ & $SU\left(N_{f}\right)$ & $SU\left(2\right)\times SU\left(2\right)$\tabularnewline
\hline 
\hline 
$X^{M}$ & $M=1,2,3,4$ & adj & 1 & (2,2)\tabularnewline
\hline 
$\omega_{i\dot{\alpha}}$ & $\dot{\alpha}=1,2$;$i=1,2...N_{f}$ & adj & fund & (1,2)\tabularnewline
\hline 
$A_{0}$ &  & adj & 1 & (1,1)\tabularnewline
\hline 
$D_{s}$ & $s=1,2,3$ & 1 & 1 & (1,3)\tabularnewline
\hline 
\end{tabular}
\par\end{centering}

\caption{Fields in the matrix model}
\end{table}

\noindent $A_{0}$ and $\vec{D}$ are auxiliary fields while $X$
and $\omega$ are dynamical fields. For a more realistic case and
simplicity, only the two-flavor case is considered throughout this
paper, i.e. $N_{f}=2$. In the action (\ref{eq:Matrix action}), the
trace is taken over the indices of $U\left(k\right)$ group. The total
symmetry of the matrix model (\ref{eq:Matrix action}) is $U\left(k\right)\times SU\left(N_{f}\right)\times SO\left(3\right)$,
where the first $U\left(k\right)$ group is a local symmetry group
while the last $SO\left(3\right)$ represents the spatial rotation
group which forms a broken $SO\left(4\right)\simeq SU\left(2\right)\times SU\left(2\right)$
in the holographic dimension as shown in the table. The broken symmetry
yields the mass terms of $X^{4}$ and $\omega$. The covariant derivatives
in action (\ref{eq:Matrix action}) are defined as $D_{0}X^{M}=\partial_{0}X^{M}-i\left[A_{0},X^{M}\right]$,
$D_{0}\omega=\partial_{0}\omega-iA_{0}\omega$ and $D_{0}\bar{\omega}=\partial_{0}\bar{\omega}+iA_{0}\bar{\omega}$.
The indices of spinor for $X$ are defined as $X_{\alpha\dot{\alpha}}=\left(X^{M}\sigma_{M}\right)_{\alpha\dot{\alpha}}$
and $\sigma_{M}=\left(i\vec{\tau},1\right)$, $\bar{\sigma}_{M}=\left(-i\vec{\tau},1\right)$
where $\vec{\tau}$'s are Pauli matrices since only two-flavor case
($N_{f}=2$) is the concern. Other details about this matrix model
from the Sakai-Sugimoto model in the D0-D4 background are in \cite{key-08}.

\subsection{Two-body effective force for baryons from D0-D4 matrix model}

Let us explain briefly how to calculate the two-body effective force
for baryons at short distances from the matrix model (\ref{eq:Matrix action})
(See \cite{key-08} for the complete review), and it is also a parallel
computation to \cite{key-12}. We first obtain the two-body Hamiltonian
by integrating out the auxiliary field $A_{0}$ and describe a single
baryon by its wave function. However, the key here is to solve the
``ADHM constraint'' \cite{key-21} to minimize the potential introduced
after integrating out the other auxiliary field $\vec{D}$. Since
only two-flavor case ($N_{f}=2$) is the concern, the ADHM constraint
could be written exactly as

\begin{equation}
\vec{\tau}_{\ \dot{\beta}}^{\dot{\alpha}}\left(\bar{X}^{\dot{\beta}\alpha}X_{\dot{\alpha}\alpha}+\bar{\omega}^{\dot{\beta}\alpha}\omega_{\dot{\alpha}\alpha}\right)_{BA}=0\ ,\label{eq:ADHM constraint}
\end{equation}

\noindent with the indices $A,B=1,2...k$.

The equation (\ref{eq:ADHM constraint}) could be solved by chosen
$\omega_{\dot{\alpha}i}=U_{\dot{\alpha}i}\rho$ for the case of a
single baryon, where $U$ is a $SU\left(2\right)$ matrix. For the
two-body case (i.e. $k=2$), the generic solution could be chosen
as the ADHM data of $SU\left(2\right)$ Yang-Mills instantons, which
are

\begin{equation}
X^{M}=\tau^{3}\frac{r_{M}}{2}+\tau^{1}Y_{M},\ \omega_{\dot{\alpha}i}^{A=1}=U_{\dot{\alpha}i}^{A=1}\rho_{1},\ \omega_{\dot{\alpha}i}^{A=2}=U_{\dot{\alpha}i}^{A=2}\rho_{2}\ ,\label{eq:two body ADHM data}
\end{equation}

\noindent where $Y^{M}$ is the off-diagonal part of $X^{M}$ which
is defined as

\begin{equation}
Y_{M}=-\frac{\rho_{1}\rho_{2}}{4\left(r_{L}\right)^{2}}\mathrm{Tr}\left[\bar{\sigma}_{M}r_{N}\sigma_{N}\left(\left(U^{1}\right)^{\dagger}U^{2}-\left(U^{2}\right)^{\dagger}U^{1}\right)\right]\ .\label{eq:ADHM data Y_M}
\end{equation}

\noindent We define $\left|r\right|^{2}=\left(r^{M}\right)^{2}$ and
the vector $r^{M}$ is interpreted as the distance between the two
baryons. $U^{(1)}$ and $U^{(2)}$ are all $SU\left(2\right)$ matrices
as the moduli parameters for each instanton while $\rho_{1}$ and
$\rho_{2}$ are associated with the size of each instantons. The ADHM
constraint is satisfied with this choice and the potential associated
with $\vec{D}$ in the action (\ref{eq:Matrix action}) vanishes.

With the decomposition of $U\left(2\right)\simeq U\left(1\right)\times SU\left(2\right)$,
i.e. $A_{0}=A_{0}^{0}\boldsymbol{1}_{2\times2}+A_{0}^{1}\tau^{1}+A_{0}^{2}\tau^{2}+A_{0}^{3}\tau^{3}$,
it is straightforward to obtain the two-body Hamiltonian after integrating
out the auxiliary field $A_{0}$ to evaluate the terms with $A_{0}$
in the action (\ref{eq:Matrix action}),

\begin{eqnarray}
S_{kinetic+CS}^{\mathrm{on-shell}} & = & \frac{\lambda N_{c}M_{KK}}{54\pi}\left(1+\zeta\right)^{3/2}\mathrm{Tr}\int dt\left[\left(D_{0}X^{M}\right)^{2}+D_{0}\bar{\omega}_{i}^{\dot{\alpha}}D_{0}\omega_{i\dot{\alpha}}\right]+N_{c}\mathrm{Tr}\int dtA_{0}\nonumber \\
 & = & \frac{\lambda N_{c}M_{KK}}{54\pi}\left(1+\zeta\right)^{3/2}\int dt\bigg\{2\left(A_{0}^{1}\right)^{2}r_{M}^{2}+8\left(A_{0}^{3}\right)^{2}Y_{M}^{2}\nonumber \\
 &  & +2\left(\rho_{1}^{2}+\rho_{2}^{2}\right)\left[\left(A_{0}^{0}\right)^{2}+\left(A_{0}^{1}\right)^{2}+\left(A_{0}^{3}\right)^{2}\right]\nonumber \\
 &  & +4\rho_{1}\rho_{2}A_{0}^{0}A_{0}^{1}\mathrm{Tr}\left[\left(U^{1}\right)^{\dagger}U^{2}\right]+4\left(\rho_{1}^{2}-\rho_{2}^{2}\right)A_{0}^{0}A_{0}^{3}+\frac{108\pi}{\lambda M_{KK}}\left(1+\zeta\right)^{-3/2}A_{0}\bigg\}\ .\label{eq:kinetic onshell}
\end{eqnarray}
We need to substitute the solutions for all the components of $A_{0}$
back into (\ref{eq:kinetic onshell}) once we solve the equations
of motion for $A_{0}$. The potential could be evaluated by using
$\int dtV=-S_{\mathrm{on-shell}}$ as

\begin{eqnarray}
V & = & 2V_{1-\mathrm{body}}+V_{2-\mathrm{body}},\ \ V_{1-\mathrm{body}}\ =\ \frac{27\pi N_{c}}{4\lambda M_{KK}}\frac{1}{\left(1+\zeta\right)^{3/2}},\nonumber \\
V_{2-\mathrm{body}} & = & \frac{27\pi N_{c}}{\lambda M_{KK}}\frac{1}{\left(1+\zeta\right)^{3/2}}\frac{u_{0}^{2}}{\left|r\right|^{2}+2\rho^{2}-2u_{0}^{2}\rho^{2}}.\label{eq:1,2 body force}
\end{eqnarray}

\noindent Here we have used the same notation as \cite{key-19} by
defining $u_{0}=\frac{1}{2}\left(\mathrm{Tr}\left[\left(U^{1}\right)^{\dagger}U^{2}\right]\right)$
with the choice of $\rho_{1}=\rho_{2}=\rho$ and kept the leading
term in the large $N_{c}$ expansion only. 

There is also an additional term to (\ref{eq:kinetic onshell}) which
is the mass term for $X^{4}$ in the action (\ref{eq:Matrix action}), 

\begin{eqnarray}
 & \frac{\lambda N_{c}M_{KK}}{54\pi}\left(1+\zeta\right)^{3/2}\frac{2}{3}\left(1-\frac{1}{2}\zeta\right)M_{KK}^{2}\mathrm{Tr}\left(X^{4}\right)^{2}\nonumber \\
= & \frac{\lambda N_{c}M_{KK}}{81\pi}\left(1+\zeta\right)^{3/2}\left(1-\frac{1}{2}\zeta\right)M_{KK}^{2}\left(\frac{r_{4}^{2}}{2}+Y_{4}^{2}\right) & .\label{eq:mass term X4}
\end{eqnarray}

\noindent Thus there is an additional two-body potential from the
off-diagonal components of $Y$ which is

\begin{equation}
V_{2-\mathrm{body}}^{\mathrm{mass}}=\frac{\lambda N_{c}M_{KK}}{162\pi}\left(1+\zeta\right)^{3/2}\left(1-\frac{1}{2}\zeta\right)M_{KK}^{2}\left[\frac{\rho_{1}^{2}\rho_{2}^{2}}{\left(r_{M}^{2}\right)^{2}}\left(r_{i}\mathrm{Tr}\left[i\tau^{i}U^{(1)\dagger}U^{(2)}\right]\right)^{2}\right]\label{eq:2-body mass}
\end{equation}

\noindent with $i=1,2,3$. So we have the total two-body potential
which is the sum of (\ref{eq:1,2 body force}) and (\ref{eq:2-body mass}).
Note that the four-dimensional inter-baryon distance $\left|r\right|^{2}$
is equal to the distance between baryons in three dimensions, since
for the leading order in the large $N_{c}$ expansion, the classical
value of the $X^{4}$ vanishes for the single instantons.

Finally, in order to evaluate the vacuum expectation of the potential
(\ref{eq:1,2 body force}) and (\ref{eq:2-body mass}), we need to
use the nucleon wave function as in \cite{key-08,key-12,key-19,key-22},
which is

\begin{equation}
\frac{1}{\pi}\left(\tau^{2}U\right)_{IJ}=\left(\begin{array}{cc}
|p\uparrow> & |p\downarrow>\\
|n\uparrow> & |n\downarrow>
\end{array}\right)_{IJ}=\frac{1}{\pi}\left(\begin{array}{cc}
a_{1}+ia_{2} & -a_{3}-ia_{4}\\
-a_{3}+ia_{4} & -a_{1}+ia_{2}
\end{array}\right)_{IJ}.\label{eq:wave function for nucleon}
\end{equation}
The $SU\left(2\right)$ matrix $U$ represents a unit 4-vector as
$U=ia_{i}\tau^{i}+a_{4}\boldsymbol{1}_{2\times2}$ with the normalization
$\left(a_{1}\right)^{2}+\left(a_{2}\right)^{2}+\left(a_{3}\right)^{2}+\left(a_{4}\right)^{2}=1$.
Using the standard definition $S_{12}=12J_{1}^{i}\hat{r}^{i}J_{2}^{j}\hat{r}^{j}-4J_{1}^{i}J_{2}^{i}$
with $\hat{r}^{i}=r^{i}/\left|r\right|$ and $i=1,2,3$, it yields
the form $\left\langle V\right\rangle _{I_{1},I_{2},J_{1},J_{2}}=V_{C}\left(\vec{r}\right)+S_{12}V_{T}\left(\vec{r}\right)$
as the potential of two -body nucleons. Then we obtain a central and
a tensor part of the two-body force at short distances which are

\begin{eqnarray}
V_{C}^{(0)}\left(\vec{r}\right) & = & \pi\left[\frac{3^{3}}{2}+8\left(\vec{I}_{1}\cdot\vec{I}_{2}\right)\left(\vec{J}_{1}\cdot\vec{J}_{2}\right)\right]\frac{N_{c}}{\lambda M_{KK}}\frac{1}{\left(1+\zeta\right)^{3/2}}\frac{1}{r^{2}},\nonumber \\
V_{T}^{(0)}\left(\vec{r}\right) & = & 2\pi\left(\vec{I}_{1}\cdot\vec{I}_{2}\right)\frac{N_{c}}{\lambda M_{KK}}\frac{1}{\left(1+\zeta\right)^{3/2}}\frac{1}{r^{2}}.\label{eq:Nuclear potential zero order}
\end{eqnarray}
(\ref{eq:Nuclear potential zero order}) is the leading order term
from (\ref{eq:kinetic onshell}) in the expansion by assuming $r_{M}\gg\rho$.
And we also have the next to the leading order terms in \cite{key-08}
which are

\begin{eqnarray}
V_{C}^{(1)}\left(\vec{r}\right) & = & \left[\frac{1}{81}-\frac{16}{2187}\left(\vec{I}_{1}\cdot\vec{I}_{2}\right)\left(\vec{J}_{1}\cdot\vec{J}_{2}\right)\right]\left(1+\zeta\right)^{3/2}\left(1-\frac{1}{2}\zeta\right)\frac{\lambda N_{c}M_{KK}^{3}}{\pi}\frac{\rho^{4}}{r^{2}},\nonumber \\
V_{T}^{(1)}\left(\vec{r}\right) & = & \frac{8}{2187}\left(\vec{I}_{1}\cdot\vec{I}_{2}\right)\left(1+\zeta\right)^{3/2}\left(1-\frac{1}{2}\zeta\right)\frac{\lambda N_{c}M_{KK}^{3}}{\pi}\frac{\rho^{4}}{r^{2}}.\label{eq:Nuclear potential leading order}
\end{eqnarray}

\noindent As we can see, the two-body potential is modified by the
appearance of smeared D0-branes, i.e. considering the non-trivial
QCD vacuum. And there would be an attractive force in (\ref{eq:Nuclear potential leading order})
if $\zeta>2$. Thus it is just the constraint for stable baryonic
states in the D0-D4/D8 system in two-body case, which is the same
as the constraint for the state of the single baryonic state and is
quite different from the original Sakai-Sugimoto model obviously.

\section{Three-body force for baryons at short distances}

In this section we will test our matrix model (\ref{eq:Matrix action})
by computing the three-body effective potential (i.e. $k=3$) at short
distances. It is a parallel procedure to the case of two-body potential
for baryons as in the previous section, and we follow the same procedures
as mentioned in \cite{key-19} which are
\begin{description}
\item [{A.}] Choose the value of $k$ (the number of baryons), and solve
the ADHM constraint (i.e. integrate out the auxiliary field $\vec{D}$
and minimize the ADHM potential).
\item [{B.}] Substitute the solution back into the action (\ref{eq:Matrix action})
of the matrix model.
\item [{C.}] Integrate out the auxiliary field $A_{0}$.
\item [{D.}] Evaluate the Hamiltonian with the desired baryonic state.
\end{description}
As a first test, we will consider a case that all three baryons take
the same classical spin or isospin. And secondly, we will demonstrate
the computations for baryons which are aligned on a straight line
with equal spacings after taking the explicit set-up for generic quantum
spin or isospin to our system. In fact we are also less clear about
how to get a physical interpretation from the calculations for the
baryons with generic positions, thus we also choose this linear position
to simplify and illuminate our calculations. Finally, we finish our
tests by evaluating the three-body Hamiltonian with two specific three-body
quantum states. They are the states of three neutrons with averaged
spins and proton-proton-neutron (or proton-neutron-proton).

\subsection{Three-body force for baryons with classical spin/isospin}

Our goal is to evaluate the three-body Hamiltonian, so let us start
with the four steps mentioned above.

\subsubsection*{A. Solve the ADHM constraint}

First, we need to consider minimizing the ADHM potential in the matrix
model. This is equivalent to solve the ADHM constraint for any $A,B=1,2,3$.
A simple solution to this constraint is

\begin{eqnarray}
\omega_{\dot{\alpha}i}^{A} & = & U_{\dot{\alpha}i}\rho^{A},\ \ \ \left(A=1,2,3\right)\nonumber \\
X^{M} & = & \sum_{a=3,8}\frac{\lambda^{a}}{2}r_{a}^{M}.\label{eq:ADHM data for cls}
\end{eqnarray}

\noindent (\ref{eq:ADHM data for cls}) is nothing but the ADHM data
for the 't Hooft instantons which has been used in \cite{key-08,key-12,key-19}.
Noting that the $2\times2$ unitary matrix $U$ does not depend on
the index $A$ while the degrees of freedom $\omega$ correspond to
the spin and isospin. The matrices $X$ are diagonal and their diagonal
elements represent the location of the baryons with $M=1,2,3$. The
$X$'s and special $\omega$'s of the ADHM data for the 't Hooft instantons
(\ref{eq:ADHM data for cls}) are sharing the same orientation. ``Classical
spin/isospin'' here means that in fact we can not fix the orientation
$U$ and consider the wave functions with finite width at same time.
All the terms with the commutators $\left[X,X\right]$ vanish since
the matrices $X$ are diagonal, which obviously satisfies the ADHM
constraint.

\subsubsection*{B. Substitute the ADHM data to the matrix action}

The inter-baryon potential comes from the terms after integrating
out the auxiliary field $A_{0}$ and the mass term of $X^{4}$. However
the mass term of $X^{4}$ vanishes for the 't Hooft instantons because
of no off-diagonal components in (\ref{eq:ADHM data for cls}) . And
on the other hand, we are going to choose the gauge $\partial_{0}\omega_{\dot{\alpha}i}^{A}=0$,
thus there is no time dependence in $\omega$ or $X$. So in this
section, we need to consider the terms related to $A_{0}$ only. 

With the gauge group $U\left(k\right)$ and $k=3$ for three-body
case, the auxiliary field $A_{0}$ could be written exactly by the
Gell-Mann matrices $\lambda^{a}$ which is 

\begin{equation}
A_{0}=A_{0}^{0}\boldsymbol{1}_{3\times3}+\sum_{a=1}^{8}A_{0}^{a}\frac{\lambda^{a}}{2}.\label{eq:A0 field}
\end{equation}

\noindent As in the two-body case, only the kinetic terms of $\omega$'s
and $X$'s contain $A_{0}$ while only $U\left(1\right)$ component
$A_{0}^{0}$ appears in the CS term in action (\ref{eq:Matrix action}).
So for the kinetic term of $X$, it takes the following exact forms
by substituting the ADHM data (\ref{eq:ADHM data for cls}),

\begin{eqnarray}
\mathrm{Tr}\left(D_{0}X^{M}\right)^{2} & = & \frac{1}{2}\left[\left(A_{0}^{1}r_{3}^{M}\right)^{2}+\left(A_{0}^{2}r_{3}^{M}\right)^{2}\right]+\frac{1}{8}\left[\left(A_{0}^{4}\right)^{2}+\left(A_{0}^{5}\right)^{2}\right]\left(r_{3}^{M}+\sqrt{3}r_{8}^{M}\right)^{2}\nonumber \\
 &  & +\frac{1}{8}\left[\left(A_{0}^{6}\right)^{2}+\left(A_{0}^{7}\right)^{2}\right]\left(r_{3}^{M}-\sqrt{3}r_{8}^{M}\right)^{2}.\label{eq:kinetic term X}
\end{eqnarray}
And then we need to consider the kinetic term for $\omega$. It is
a lengthy but straightforward calculation just by using the ADHM data
(\ref{eq:ADHM data for cls}). As a result, we obtain the following
expression,

\begin{eqnarray}
\mathrm{Tr}\left(D_{0}\bar{\omega}_{i}^{\dot{\alpha}}D_{0}\omega_{i\dot{\alpha}}\right) & = & 2\left[\left(\rho^{1}\right)^{2}+\left(\rho^{2}\right)^{2}+\left(\rho^{3}\right)^{2}\right]\left[\left(A_{0}^{0}\right)^{2}+\frac{1}{6}\sum_{a=1}^{8}\left(A_{0}^{a}\right)^{2}\right]+4\rho^{1}\rho^{2}A_{0}^{1}A_{0}^{0}+4\rho^{1}\rho^{3}A_{0}^{4}A_{0}^{0}\nonumber \\
 &  & +4\rho^{2}\rho^{3}A_{0}^{6}A_{0}^{0}+2A_{0}^{3}A_{0}^{0}\left[\left(\rho^{1}\right)^{2}-\left(\rho^{2}\right)^{2}\right]+\frac{2}{\sqrt{3}}A_{0}^{8}A_{0}^{0}\left[\left(\rho^{1}\right)^{2}+\left(\rho^{2}\right)^{2}-2\left(\rho^{3}\right)^{2}\right]\nonumber \\
 &  & +\frac{2\rho^{1}\rho^{2}}{\sqrt{3}}A_{0}^{1}A_{0}^{8}+\rho^{1}\rho^{2}A_{0}^{4}A_{0}^{6}+\rho^{1}\rho^{2}A_{0}^{5}A_{0}^{7}-\frac{\rho^{1}\rho^{3}}{\sqrt{3}}A_{0}^{4}A_{0}^{8}+\rho^{1}\rho^{3}A_{0}^{1}A_{0}^{6}-\rho^{1}\rho^{3}A_{0}^{2}A_{0}^{7}\nonumber \\
 &  & +\rho^{1}\rho^{3}A_{0}^{3}A_{0}^{4}-\frac{\rho^{2}\rho^{3}}{\sqrt{3}}A_{0}^{6}A_{0}^{8}+\rho^{2}\rho^{3}A_{0}^{1}A_{0}^{4}+\rho^{2}\rho^{3}A_{0}^{2}A_{0}^{5}-\rho^{2}\rho^{3}A_{0}^{3}A_{0}^{6}\nonumber \\
 &  & +\left[\frac{1}{\sqrt{3}}A_{0}^{3}A_{0}^{8}+\frac{1}{4}\left(A_{0}^{4}\right)^{2}+\frac{1}{4}\left(A_{0}^{5}\right)^{2}-\frac{1}{4}\left(A_{0}^{6}\right)^{2}-\frac{1}{4}\left(A_{0}^{7}\right)^{2}\right]\left[\left(\rho^{1}\right)^{2}-\left(\rho^{2}\right)^{2}\right]\nonumber \\
 &  & +\frac{1}{12}\left[2\left(A_{0}^{1}\right)^{2}+2\left(A_{0}^{2}\right)^{2}+2\left(A_{0}^{3}\right)^{2}-2\left(A_{0}^{8}\right)^{2}-\left(A_{0}^{4}\right)^{2}-\left(A_{0}^{5}\right)^{2}-\left(A_{0}^{6}\right)^{2}-\left(A_{0}^{7}\right)^{2}\right]\nonumber \\
 &  & \times\left[\left(\rho^{1}\right)^{2}+\left(\rho^{2}\right)^{2}-2\left(\rho^{3}\right)^{2}\right].\label{eq:kinetic term w}
\end{eqnarray}
In fact (\ref{eq:kinetic term X}) and (\ref{eq:kinetic term w})
are as same as the ``data'' used in \cite{key-19} since we start
with the same ADHM data (\ref{eq:ADHM data for cls}) for the 't Hooft
instantons.

And finally, we also need to consider the Lagrangian for CS term which
is

\begin{equation}
L_{CS}=\frac{162\pi}{\lambda M_{KK}\left(1+\zeta\right)^{3/2}}A_{0}^{0}.\label{eq:CS term L-1}
\end{equation}
So the total Lagrangian involving the field $A_{0}$ consists of (\ref{eq:kinetic term X})
(\ref{eq:kinetic term w}) and (\ref{eq:CS term L-1}), which is denoted
by $L_{A_{0}}$, 

\begin{equation}
L_{A_{0}}=\mathrm{Tr}\left(D_{0}X^{M}\right)^{2}+\mathrm{Tr}\left(D_{0}\bar{\omega}_{i}^{\dot{\alpha}}D_{0}\omega_{i\dot{\alpha}}\right)+L_{CS}.\label{eq:LA0-1}
\end{equation}

\subsubsection*{C. Integrate out the auxiliary field $A_{0}$}

We need to solve all the equations of motion for all the components
of $A_{0}$. All the equations of motion can be obtained by Euler-Lagrange
equation with (\ref{eq:LA0-1}), i.e.

\begin{equation}
\frac{\partial L_{A_{0}}}{\partial A_{0}^{0}}=0,\ \ \mathrm{or}\ \ \frac{\partial L_{A_{0}}}{\partial A_{0}^{a}}=0\ \ \ \left(\mathrm{for}\ \ a=1,2...8\right).\label{eq:EOM for A}
\end{equation}
We obtain 9 equations while all the components of $A_{0}$ are mixed
to each other. By solving these equations we find a unique solution
as\footnote{In order to simplified the formula, we have used $r_{a}^{2}$ to represent
$r_{a}^{M}r_{a}^{M}$ and $r_{a}r_{b}$ to represent $r_{a}^{M}r_{b}^{M}$.}

\begin{eqnarray}
A_{0}^{0} & = & \frac{9\pi}{2\left(1+\zeta\right)^{3/2}\lambda\left(\rho^{1}\right)^{2}\left(\rho^{2}\right)^{2}\left(\rho^{3}\right)^{2}M_{KK}r_{3}^{2}\left(r_{3}^{2}-3r_{8}^{2}\right)^{2}}\times\nonumber \\
 &  & \bigg\{-r_{3}^{2}\left(\rho^{1}\right)^{2}\left(\rho^{2}\right)^{2}\bigg[9\left(r_{8}^{2}\right)^{2}+8\sqrt{3}r_{3}r_{8}\left(\left(\rho^{2}\right)^{2}-\left(\rho^{1}\right)^{2}\right)\nonumber \\
 &  & -6r_{8}^{2}\left(r_{3}^{2}-2\left(\left(\rho^{2}\right)^{2}+\left(\rho^{1}\right)^{2}\right)\right)+r_{3}^{2}\left(r_{3}^{2}+4\left(\left(\rho^{2}\right)^{2}+\left(\rho^{1}\right)^{2}\right)\right)\bigg]\nonumber \\
 &  & -\bigg[\left(r_{3}^{2}-3r_{8}^{2}\right)^{2}\left(\rho^{1}\right)^{2}\left(r_{3}^{2}+\left(\rho^{1}\right)^{2}\right)+\left(r_{3}^{2}-3r_{8}^{2}\right)^{2}\left(\rho^{2}\right)^{4}\nonumber \\
 &  & +\left(\left(r_{3}^{2}-3r_{3}r_{8}\right)^{2}r_{3}^{2}+18\left(r_{3}^{2}+r_{8}^{2}\right)\left(\rho^{1}\right)^{2}\right)\left(\rho^{2}\right)^{2}\bigg]\left(\rho^{3}\right)^{2}\nonumber \\
 &  & +4\left[2\sqrt{3}r_{3}r_{8}\left(\left(\rho^{2}\right)^{2}-\left(\rho^{1}\right)^{2}\right)-r_{3}^{2}\left(\left(\rho^{2}\right)^{2}+\left(\rho^{1}\right)^{2}\right)-3r_{8}^{2}\left(\left(\rho^{2}\right)^{2}+\left(\rho^{1}\right)^{2}\right)\right]\nonumber \\
 &  & \times r_{3}^{2}\left(\rho^{3}\right)^{4}\bigg\},\nonumber \\
A_{0}^{3} & = & \frac{27\pi}{2\left(1+\zeta\right)^{3/2}\lambda\left(\rho^{1}\right)^{2}\left(\rho^{2}\right)^{2}M_{KK}r_{3}^{2}\left(r_{3}^{2}-3r_{8}^{2}\right)^{2}}\times\nonumber \\
 &  & \bigg\{\left(r_{3}^{2}\right)^{3}\left[\left(\rho^{1}\right)^{2}-\left(\rho^{2}\right)^{2}\right]+9\left(r_{8}^{2}\right)^{2}\left[\left(\rho^{1}\right)^{4}-\left(\rho^{2}\right)^{4}\right]+\left(r_{3}^{2}\right)^{2}\left[\left(\rho^{1}\right)^{2}-\left(\rho^{2}\right)^{2}\right]\nonumber \\
 &  & \times\left(\left(\rho^{1}\right)^{2}+\left(\rho^{2}\right)^{2}+4\left(\rho^{3}\right)^{2}-6r_{8}^{2}\right)+\left[2\left(\rho^{1}\right)^{2}\left(\rho^{2}\right)^{2}+\left(\left(\rho^{1}\right)^{2}+\left(\rho^{2}\right)^{2}\right)\left(\rho^{3}\right)^{2}\right]\nonumber \\
 &  & \times8\sqrt{3}r_{3}^{2}\left(r_{3}r_{8}\right)+3r_{3}^{2}r_{8}^{2}\left[\left(\rho^{1}\right)^{2}-\left(\rho^{2}\right)^{2}\right]\times\nonumber \\
 &  & \left[3r_{8}^{2}-2\left(\left(\rho^{1}\right)^{2}+\left(\rho^{2}\right)^{2}-2\left(\rho^{3}\right)^{2}\right)\right]\bigg\},\nonumber \\
A_{0}^{8} & = & \frac{9\pi}{2\left(1+\zeta\right)^{3/2}\lambda\left(\rho^{1}\right)^{2}\left(\rho^{2}\right)^{2}\left(\rho^{3}\right)^{2}M_{KK}r_{3}^{2}\left(r_{3}^{2}-3r_{8}^{2}\right)^{2}}\times\nonumber \\
 &  & \bigg\{-9\sqrt{3}\left(r_{8}^{2}\right)^{2}\left[\left(\rho^{1}\right)^{2}+\left(\rho^{2}\right)^{2}\right]^{2}\left(\rho^{3}\right)^{2}+\left[2\left(\rho^{1}\right)^{2}\left(\rho^{2}\right)^{2}-\left(\left(\rho^{1}\right)^{2}+\left(\rho^{2}\right)^{2}\right)\left(\rho^{3}\right)^{2}\right]\nonumber \\
 &  & \times\sqrt{3}\left(r_{3}^{2}\right)^{3}-24r_{3}^{2}\left(r_{3}r_{8}\right)\left[\left(\rho^{1}\right)^{2}-\left(\rho^{2}\right)^{2}\right]\left[2\left(\rho^{1}\right)^{2}\left(\rho^{2}\right)^{2}+\left(\rho^{3}\right)^{4}\right]\nonumber \\
 &  & +3\sqrt{3}r_{3}^{2}r_{8}^{2}\bigg[2\left(\rho^{1}\right)^{2}\left(\rho^{2}\right)^{2}\left(3r_{8}^{2}+4\left(\rho^{1}\right)^{2}+4\left(\rho^{2}\right)^{2}\right)-4\left(\left(\rho^{1}\right)^{2}+\left(\rho^{2}\right)^{2}\right)\left(\rho^{3}\right)^{4}\nonumber \\
 &  & -3r_{8}^{2}\left(\rho^{3}\right)^{2}\left(\left(\rho^{1}\right)^{2}+\left(\rho^{2}\right)^{2}\right)+2\left(\left(\rho^{1}\right)^{4}+\left(\rho^{2}\right)^{4}+6\left(\rho^{1}\right)^{2}\left(\rho^{2}\right)^{2}\right)\left(\rho^{3}\right)^{2}\bigg]\nonumber \\
 &  & -\sqrt{3}\left(r_{3}^{2}\right)^{2}\bigg[-4\left(\rho^{1}\right)^{2}\left(\rho^{2}\right)^{2}\left(-3r_{8}^{2}+2\left(\rho^{1}\right)^{2}+2\left(\rho^{2}\right)^{2}\right)+\left(\rho^{3}\right)^{2}\times\nonumber \\
 &  & \left(\left(\rho^{1}\right)^{4}-6\left(\rho^{1}\right)^{2}\left(\rho^{2}\right)^{2}-6r_{8}^{2}\left(\rho^{1}\right)^{2}-6r_{8}^{2}\left(\rho^{2}\right)^{2}\right)+4\left(\rho^{3}\right)^{2}\left(\left(\rho^{1}\right)^{2}+\left(\rho^{2}\right)^{2}\right)\text{\ensuremath{\bigg]}}\bigg\},\label{eq:solution for A - 1}
\end{eqnarray}
and

\begin{align}
A_{0}^{1} & =\ \frac{27\pi\left[\left(\rho^{2}\right)^{2}+\left(\rho^{1}\right)^{2}\right]}{\left(1+\zeta\right)^{3/2}\lambda M_{KK}\rho^{1}\rho^{2}r_{3}^{2}},\nonumber \\
A_{0}^{4} & =\ \frac{108\sqrt{3}\pi\left[\left(\rho^{1}\right)^{2}+\left(\rho^{3}\right)^{2}\right]}{\left(1+\zeta\right)^{3/2}\lambda M_{KK}\rho^{1}\rho^{3}\left[\sqrt{3}r_{3}^{2}+6r_{3}r_{8}+3\sqrt{3}r_{8}^{2}\right]},\nonumber \\
A_{0}^{6} & =\frac{108\pi\left[r_{3}^{2}+2\sqrt{3}r_{3}r_{8}+3r_{8}^{2}\right]\left[\left(\rho^{2}\right)^{2}+\left(\rho^{3}\right)^{2}\right]}{\left(1+\zeta\right)^{3/2}\lambda M_{KK}\rho^{2}\rho^{3}\left(r_{3}^{2}-3r_{8}^{2}\right)^{2}},\label{eq:solution for A - 2}
\end{align}

\noindent while the solutions for other components are $A_{0}^{2}=A_{0}^{5}=A_{0}^{7}=0$.
We can obtain the integrated Lagrangian from $L_{A_{0}}$ with the
moduli parameters $r_{3}^{M},\ r_{8}^{M}$ and $\rho^{A}$ by plugging
the solutions (\ref{eq:solution for A - 1}) and (\ref{eq:solution for A - 2})
back to (\ref{eq:LA0-1}), which is

\begin{eqnarray}
L_{A_{0}}\left(r_{3}^{M},r_{8}^{M},\rho^{A}\right) & = & -\left(\frac{54\pi}{\lambda M_{KK}}\right)^{2}\frac{1}{\left(1+\zeta\right)^{3}}\bigg[\frac{1}{8}\sum_{A=1}^{3}\frac{1}{\left(\rho^{A}\right)}+\frac{1}{4\left(r_{3}^{M}\right)^{2}}\left(1+\frac{\left(\rho^{1}\right)^{2}}{2\left(\rho^{2}\right)^{2}}+\frac{\left(\rho^{2}\right)^{2}}{2\left(\rho^{1}\right)^{2}}\right)\nonumber \\
 &  & +\frac{1}{\left(r_{3}^{M}+\sqrt{3}r_{8}^{M}\right)^{2}}\left(1+\frac{\left(\rho^{1}\right)^{2}}{2\left(\rho^{3}\right)^{2}}+\frac{\left(\rho^{3}\right)^{2}}{2\left(\rho^{1}\right)^{2}}\right)\nonumber \\
 &  & +\frac{1}{\left(r_{3}^{M}-\sqrt{3}r_{8}^{M}\right)^{2}}\left(1+\frac{\left(\rho^{2}\right)^{2}}{2\left(\rho^{3}\right)^{2}}+\frac{\left(\rho^{3}\right)^{2}}{2\left(\rho^{2}\right)^{2}}\right)\bigg].\label{eq:LA onshell}
\end{eqnarray}

\noindent By employing the picture of soliton, we obtain the potential
$V^{\mathrm{cl}}$ from the D0-D4/D8 matrix model, which is

\begin{equation}
S=\frac{\lambda N_{c}M_{KK}}{54\pi}\left(1+\zeta\right)^{3/2}\int dtL_{A_{0}}=-\int dtV^{\mathrm{cl}}.\label{eq:soliton}
\end{equation}

\noindent Thus we have

\begin{eqnarray}
V^{\mathrm{cl}} & = & \frac{54\pi N_{c}}{\lambda M_{KK}}\frac{1}{\left(1+\zeta\right)^{3/2}}\bigg[\frac{1}{8}\sum_{A=1}^{3}\frac{1}{\left(\rho^{A}\right)}+\frac{1}{4\left(r_{3}^{M}\right)^{2}}\left(1+\frac{\left(\rho^{1}\right)^{2}}{2\left(\rho^{2}\right)^{2}}+\frac{\left(\rho^{2}\right)^{2}}{2\left(\rho^{1}\right)^{2}}\right)\nonumber \\
 &  & +\frac{1}{\left(r_{3}^{M}+\sqrt{3}r_{8}^{M}\right)^{2}}\left(1+\frac{\left(\rho^{1}\right)^{2}}{2\left(\rho^{3}\right)^{2}}+\frac{\left(\rho^{3}\right)^{2}}{2\left(\rho^{1}\right)^{2}}\right)\nonumber \\
 &  & +\frac{1}{\left(r_{3}^{M}-\sqrt{3}r_{8}^{M}\right)^{2}}\left(1+\frac{\left(\rho^{2}\right)^{2}}{2\left(\rho^{3}\right)^{2}}+\frac{\left(\rho^{3}\right)^{2}}{2\left(\rho^{2}\right)^{2}}\right)\bigg].\label{eq:LA onshell-1}
\end{eqnarray}
In order to obtain the potential intrinsic to the three-body case,
we have to subtract the one- and two- body Hamiltonians. It can be
read from the computation for two-body case in \cite{key-08} with
the ADHM data for the 't Hooft instantons. If we take the leading
term in the large $N_{c}$ expansion, we have the following forms

\begin{equation}
V_{1-\mathrm{body}}^{\mathrm{cl}}=\frac{27\pi N_{c}}{4\lambda M_{KK}\left(1+\zeta\right)^{3/2}}\frac{1}{\left(\rho^{A}\right)^{2}},\ \ V_{2-\mathrm{body}}^{\mathrm{cl}}=\frac{27\pi N_{c}}{4\lambda M_{KK}\left(1+\zeta\right)^{3/2}}\frac{1}{\left(r^{M}\right)^{2}}\left(2+\frac{\left(\rho^{B}\right)^{2}}{\left(\rho^{A}\right)^{2}}+\frac{\left(\rho^{A}\right)^{2}}{\left(\rho^{B}\right)^{2}}\right),\label{eq:cl 1 and 2 body}
\end{equation}
where we have used $r^{M}$ to represent the distance between the
two baryons. Then according to (\ref{eq:LA onshell-1}) and (\ref{eq:cl 1 and 2 body}),
it gives

\begin{equation}
\sum_{A=1,2,3}V_{1-\mathrm{body}}^{\left(A\right)\mathrm{cl}}+\frac{1}{2}\sum_{A\neq B}V_{2-\mathrm{body}}^{\left(A,B\right)\mathrm{cl}}=V^{\mathrm{cl}},\label{eq:25}
\end{equation}
which means the three-body force of the baryons sharing the same classical
spins or isospins vanishes exactly. The result remains as in \cite{key-19}
and in \cite{key-20} with the soliton approach.

In \cite{key-08,key-11}, we claim that the constraint for the stable
baryonic state is $\zeta<2$ in D0-D4/D8 system and it turns out the
two-body force is also affected by this constraint. However according
to (\ref{eq:LA onshell-1}) and (\ref{eq:25}), it seems this constraint
has nothing to do with our calculations for three-body case. The direct
reason is, during our calculations for the ``classical treatment''
we have set the mass term of $X^{4}$ to zero, but the constraint
for the stable baryonic state comes from this term. The physical interpretation
is, in fact we do not keep the quantum spin or isospin degrees of
freedom explicitly in this ``classical'' computation, which means
some quantum effects about the QCD vacuum in this sense is missing.
Therefore the computation based on the ``classical treatment'' is
also unfortunately unrealistic for the nucleons in the D0-D4/D8 system.
So in the next sections, we will focus on a more realistic case with
the quantum degrees of freedom for spin or isospin.

\subsection{Generic three-body force in D0-D4/D8 system}

For a generic calculation about the three-body baryons in D0-D4/D8
system, we will follow the four steps as well as the case in the previous
section.

\subsubsection*{A. Solve the ADHM constraint}

As a warm-up in the previous section, we have fixed the spins or isospins
for baryons and computed the three-body force with the ADHM data for
the 't Hooft instanton in our D0-D4/D8 matrix model, which is easy
but not realistic. In this section, we keep the quantum spin or isospin
degrees of freedom, i.e. allow arbitrary $U$ for each baryon,

\begin{equation}
\omega_{\dot{\alpha}i}^{A}=U_{\dot{\alpha}i}^{A}\rho^{A},\ \ \ \left(A=1,2,3\right).\label{eq:ADHM data w - 2}
\end{equation}

\noindent Since our D0-D4/D8 matrix model does not change the ADHM
constraint (\ref{eq:ADHM constraint}), thus we can choose the same
solution for $X$ as in \cite{key-19} which is

\begin{equation}
X^{M}=\sum_{a=3,8}\frac{\lambda^{a}}{2}r_{a}^{M}+\sum_{a=1,4,6}\frac{\lambda^{a}}{2}r_{a}^{M}.\label{eq:ADHM data X - 2}
\end{equation}
The off-diagonal components are turned on in matrices $X^{M}$ which
makes (\ref{eq:ADHM data X - 2}) different from (\ref{eq:ADHM data for cls}).
The off-diagonal $r_{1},\ r_{4}$ and $r_{6}$ should be small and
$r_{3}$ and $r_{8}$ specify the positions of these three baryons.
The classical size of baryon is small enough for large $\lambda$,
$\rho\sim\lambda^{-1/2}$, so we also need the ADHM data for well-separated
instantons since the generic three-body ADHM data is not available,
i.e.

\begin{equation}
\left|r_{3}+\sqrt{3}r_{8}\right|/2,\ \left|-r_{3}+\sqrt{3}r_{8}\right|/2,\ \left|r_{8}\right|\gg\rho^{A}.
\end{equation}

\noindent The well-separated instanton is in \cite{key-23} and we
employ it as the ADHM data in our notation, which the relevant parts
are\footnote{Here we follow the notation used in \cite{key-19,key-24}}

\begin{align}
r_{1}^{M}\sigma_{M} & =\frac{d_{12}^{M}\sigma_{M}}{\left|d_{12}\right|^{2}}\rho^{1}\rho^{2}\left[\left(U^{2}\right)^{\dagger}U^{1}-\left(U^{1}\right)^{\dagger}U^{2}\right]+\frac{\rho^{1}\rho^{2}\left(\rho^{3}\right)^{2}d_{12}^{M}\sigma_{M}}{4\left|d_{12}\right|^{2}\left|d_{13}\right|^{2}\left|d_{23}\right|^{2}}\nonumber \\
 & \times\bigg\{\left[\left(U^{3}\right)^{\dagger}U^{2}-\left(U^{2}\right)^{\dagger}U^{3}\right]d_{23}^{\dagger}d_{31}\left[\left(U^{1}\right)^{\dagger}U^{3}-\left(U^{3}\right)^{\dagger}U^{1}\right]\nonumber \\
 & -\left[\left(U^{3}\right)^{\dagger}U^{1}-\left(U^{1}\right)^{\dagger}U^{3}\right]d_{31}^{\dagger}d_{32}\left[\left(U^{2}\right)^{\dagger}U^{3}-\left(U^{3}\right)^{\dagger}U^{2}\right]\bigg\}+\mathcal{O}\left(d^{-5}\right),\nonumber \\
r_{4}^{M}\sigma_{M} & =\frac{d_{13}^{M}\sigma_{M}}{\left|d_{13}\right|^{2}}\rho^{1}\rho^{3}\left[\left(U^{3}\right)^{\dagger}U^{1}-\left(U^{1}\right)^{\dagger}U^{3}\right]+\frac{\rho^{1}\rho^{3}\left(\rho^{2}\right)^{2}d_{13}^{M}\sigma_{M}}{4\left|d_{12}\right|^{2}\left|d_{13}\right|^{2}\left|d_{23}\right|^{2}}\nonumber \\
 & \times\bigg\{\left[\left(U^{2}\right)^{\dagger}U^{3}-\left(U^{3}\right)^{\dagger}U^{2}\right]d_{23}^{\dagger}d_{21}\left[\left(U^{1}\right)^{\dagger}U^{2}-\left(U^{2}\right)^{\dagger}U^{1}\right]\nonumber \\
 & -\left[\left(U^{2}\right)^{\dagger}U^{1}-\left(U^{1}\right)^{\dagger}U^{2}\right]d_{21}^{\dagger}d_{23}\left[\left(U^{3}\right)^{\dagger}U^{2}-\left(U^{2}\right)^{\dagger}U^{3}\right]\bigg\}+\mathcal{O}\left(d^{-5}\right),\nonumber \\
r_{6}^{M}\sigma_{M} & =\frac{d_{23}^{M}\sigma_{M}}{\left|d_{23}\right|^{2}}\rho^{2}\rho^{3}\left[\left(U^{3}\right)^{\dagger}U^{2}-\left(U^{2}\right)^{\dagger}U^{3}\right]+\frac{\rho^{2}\rho^{3}\left(\rho^{1}\right)^{2}d_{23}^{M}\sigma_{M}}{4\left|d_{12}\right|^{2}\left|d_{13}\right|^{2}\left|d_{23}\right|^{2}}\nonumber \\
 & \times\bigg\{\left[\left(U^{1}\right)^{\dagger}U^{3}-\left(U^{3}\right)^{\dagger}U^{1}\right]d_{13}^{\dagger}d_{12}\left[\left(U^{2}\right)^{\dagger}U^{1}-\left(U^{1}\right)^{\dagger}U^{2}\right]\nonumber \\
 & -\left[\left(U^{1}\right)^{\dagger}U^{2}-\left(U^{2}\right)^{\dagger}U^{1}\right]d_{12}^{\dagger}d_{13}\left[\left(U^{3}\right)^{\dagger}U^{1}-\left(U^{1}\right)^{\dagger}U^{3}\right]\bigg\}+\mathcal{O}\left(d^{-5}\right),\label{eq:well-sep ADHM data}
\end{align}
with the definition of the distance vector $d_{ij}$ between the $i$-th
and $j$-th baryon,

\begin{equation}
d_{ij}=d_{ij}^{M}\sigma_{M}.\label{eq:distance vector}
\end{equation}

\noindent According to (\ref{eq:ADHM data X - 2}), we have the positions
for the three baryons respectively,

\begin{equation}
r^{M}=r_{3}^{M}/2+r_{8}^{M}/2\sqrt{3},\ -r_{3}^{M}/2+r_{8}^{M}/2\sqrt{3},\ -r_{8}^{M}/\sqrt{3},
\end{equation}
with

\begin{align}
d_{12} & =-d_{21}=r_{3},\nonumber \\
d_{13} & =-d_{31}=\left(r_{3}-\sqrt{3}r_{8}\right)/2,\nonumber \\
d_{23} & =-d_{32}=-\left(r_{3}+\sqrt{3}r_{8}\right)/2.
\end{align}

\subsubsection*{B. Substitute the ADHM data to the Lagrangian}

Here we use the $SU\left(2\right)$ matrices $U_{\dot{\alpha}i}^{A}$
to represent the rotation matrices for different three baryons with
$A=1,2,3$, which can be written as $u_{0}\mathbf{1}_{2\times2}+i\sum_{i=1}^{3}u_{i}\tau^{i}$
with $\sum_{i=1}^{3}\left(u_{i}\right)^{2}=1$. Thus the terms consists
of $U$ 's such as in (\ref{eq:well-sep ADHM data}) can be written
explicitly as

\begin{equation}
U_{\dot{\alpha}i}^{A}\left(U_{\dot{\beta}i}^{B}\right)^{\dagger}=u_{0}^{\left(AB\right)}\left(\boldsymbol{1}_{2\times2}\right)_{\dot{\alpha}\dot{\beta}}+i\sum_{i=1}^{3}u_{i}^{\left(AB\right)}\tau_{\dot{\alpha}\dot{\beta}}^{i}.\label{eq:U generic}
\end{equation}
where $u_{0}$ is defined as same as in (\ref{eq:1,2 body force}).
So we have new terms with new parameters $r_{a}^{M}$ with $a=1,4,6$
and $u_{0}^{AB}$ if compared with the ADHM data in (\ref{eq:ADHM data for cls}). 

As the case in the previous section, we also need to write the terms
including $A_{0}$, which are the kinetic terms of $X$ and $\omega$
plus the CS term. For the kinetic term of $X$, we have,

\begin{equation}
\mathrm{Tr}\left(D_{0}X^{M}\right)^{2}=\mathrm{Tr}\left(-i\left[A_{0},\sum_{a=1}^{8}\frac{\lambda^{a}}{2}r_{a}^{M}\right]\right)^{2}.\label{eq:kinetic term of X - 2}
\end{equation}
And (\ref{eq:kinetic term of X - 2}) could be simplified as

\begin{align}
\mathrm{Tr}\left(D_{0}X^{M}\right)^{2} & =\ \frac{1}{8}\bigg\{\left(A_{0}^{4}\right)^{2}r_{1}^{2}+\left(A_{0}^{6}\right)^{2}r_{1}^{2}+4\left(A_{0}^{1}\right)^{2}r_{3}^{2}+\left(A_{0}^{4}\right)^{2}r_{3}^{2}+\left(A_{0}^{6}\right)^{2}r_{3}^{2}\nonumber \\
 & \ \ \ \ -2A_{0}^{1}A_{0}^{4}r_{1}r_{4}-2\sqrt{3}A_{0}^{6}A_{0}^{8}r_{1}r_{4}+6A_{0}^{1}A_{0}^{6}r_{3}r_{4}-2\sqrt{3}A_{0}^{4}A_{0}^{8}r_{3}r_{4}\nonumber \\
 & \ \ \ \ +\left(A_{0}^{1}\right)^{2}r_{4}^{2}+\left(A_{0}^{6}\right)^{2}r_{4}^{2}+3\left(A_{0}^{8}\right)^{2}r_{4}^{2}-2A_{0}^{1}A_{0}^{6}r_{1}r_{6}-2\sqrt{3}A_{0}^{4}A_{0}^{8}r_{1}r_{6}\nonumber \\
 & \ \ \ \ -6A_{0}^{1}A_{0}^{4}r_{3}r_{6}+2\sqrt{3}A_{0}^{6}A_{0}^{8}r_{3}r_{6}-2A_{0}^{4}A_{0}^{6}r_{4}r_{6}+4\sqrt{3}A_{0}^{1}A_{0}^{8}r_{4}r_{6}\nonumber \\
 & \ \ \ \ +\left(A_{0}^{1}\right)^{2}r_{6}^{2}+\left(A_{0}^{4}\right)^{2}r_{6}^{2}+3\left(A_{0}^{8}\right)^{2}r_{6}^{2}+\left(A_{0}^{3}\right)^{2}\left[4r_{1}^{2}+r_{4}^{2}+r_{6}^{2}\right]\nonumber \\
 & \ \ \ \ +4\sqrt{3}A_{0}^{4}A_{0}^{6}r_{1}r_{8}+2\sqrt{3}\left(A_{0}^{4}\right)^{2}r_{3}r_{8}-2\sqrt{3}\left(A_{0}^{6}\right)^{2}r_{3}r_{8}-2\sqrt{3}A_{0}^{1}A_{0}^{6}r_{4}r_{8}\nonumber \\
 & \ \ \ \ -6A_{0}^{4}A_{0}^{8}r_{4}r_{8}-2\sqrt{3}A_{0}^{1}A_{0}^{4}r_{6}r_{8}-6A_{0}^{6}A_{0}^{8}r_{6}r_{8}+3\left(A_{0}^{4}\right)^{2}r_{8}^{2}+3\left(A_{0}^{6}\right)^{2}r_{8}^{2}\nonumber \\
 & \ \ \ \ -2A_{0}^{3}\bigg[4A_{0}^{1}r_{1}r_{3}+3A_{0}^{6}r_{1}r_{4}+A_{0}^{4}r_{3}r_{4}-\sqrt{3}A_{0}^{8}r_{4}^{2}-3A_{0}^{4}r_{1}r_{6}+A_{0}^{6}r_{3}r_{6}\nonumber \\
 & \ \ \ \ +\sqrt{3}A_{0}^{8}r_{6}^{2}+\sqrt{3}A_{0}^{4}r_{4}r_{8}-\sqrt{3}A_{0}^{6}r_{6}r_{8}\bigg]\bigg\}.\label{eq:kinetic X - 2}
\end{align}
Note that (\ref{eq:kinetic X - 2}) does not include the terms of
$A_{0}^{a}$ with $a=2,5,7$. We have omitted these terms since all
$A_{0}^{a}$ with $a=2,5,7$ appear in the Lagrangian as quadratic
terms which yields $A_{0}^{a=2,5,7}=0$ by their equations of motion.
According to these, we have the kinetic term for $\omega$ which is
similar to the case of the 't Hooft instanton as follow,

\begin{align}
\mathrm{Tr}\left(D_{0}\bar{\omega}_{i}^{\dot{\alpha}}D_{0}\omega_{\dot{\alpha}i}\right) & =2\left[\left(\rho^{1}\right)^{2}+\left(\rho^{2}\right)^{2}+\left(\rho^{3}\right)^{2}\right]\left[\left(A_{0}^{0}\right)^{2}+\frac{1}{6}\sum_{a=1,3,4,6,8}\left(A_{0}^{a}\right)^{2}\right]+4\rho^{1}\rho^{2}u_{0}^{(12)}A_{0}^{1}A_{0}^{0}\nonumber \\
 & \ \ +4\rho^{1}\rho^{3}u_{0}^{(13)}A_{0}^{4}A_{0}^{0}+4\rho^{2}\rho^{3}u_{0}^{(23)}A_{0}^{6}A_{0}^{0}+2A_{0}^{3}A_{0}^{0}\left[\left(\rho^{1}\right)^{2}-\left(\rho^{2}\right)^{2}\right]\nonumber \\
 & \ \ +\frac{2}{\sqrt{3}}A_{0}^{8}A_{0}^{0}\left[\left(\rho^{1}\right)^{2}+\left(\rho^{2}\right)^{2}-2\left(\rho^{3}\right)^{2}\right]+\frac{2\rho^{1}\rho^{2}u_{0}^{\left(12\right)}}{\sqrt{3}}A_{0}^{1}A_{0}^{8}+\rho^{1}\rho^{2}u_{0}^{\left(12\right)}A_{0}^{4}A_{0}^{6}\nonumber \\
 & \ \ -\frac{\rho^{1}\rho^{2}u_{0}^{\left(13\right)}}{\sqrt{3}}A_{0}^{4}A_{0}^{8}+\rho^{1}\rho^{3}u_{0}^{(13)}A_{0}^{1}A_{0}^{6}+\rho^{1}\rho^{3}u_{0}^{(13)}A_{0}^{3}A_{0}^{4}\nonumber \\
 & \ \ -\frac{\rho^{2}\rho^{3}u_{0}^{\left(23\right)}}{\sqrt{3}}A_{0}^{6}A_{0}^{8}+\rho^{2}\rho^{3}u_{0}^{(23)}A_{0}^{1}A_{0}^{4}-\rho^{2}\rho^{3}u_{0}^{(23)}A_{0}^{3}A_{0}^{6}\nonumber \\
 & \ \ +\left[\frac{1}{\sqrt{3}}A_{0}^{3}A_{0}^{8}+\frac{1}{4}\left(A_{0}^{4}\right)^{2}-\frac{1}{4}\left(A_{0}^{6}\right)^{2}\right]\left[\left(\rho^{1}\right)^{2}-\left(\rho^{2}\right)^{2}\right]\nonumber \\
 & \ \ +\frac{1}{12}\left[2\left(A_{0}^{1}\right)^{2}+2\left(A_{0}^{3}\right)^{2}-2\left(A_{0}^{8}\right)^{2}-\left(A_{0}^{4}\right)^{2}-\left(A_{0}^{6}\right)^{2}\right]\left[\left(\rho^{1}\right)^{2}+\left(\rho^{2}\right)^{2}-2\left(\rho^{3}\right)^{2}\right].\label{eq:kinetic w - 2}
\end{align}
We have used $\omega_{\dot{\alpha}i}^{A}\lambda_{AB}^{a}\left(\omega_{\dot{\alpha}i}^{B}\right)^{*}=0$
for $a=2,5,7$ since they are proportional to $U_{\dot{\alpha}i}^{A}\left(U_{\dot{\alpha}i}^{B}\right)^{\dagger}-U_{\dot{\alpha}i}^{B}\left(U_{\dot{\alpha}i}^{A}\right)^{\dagger}$
with $A,B=1,2,3$. So the total Lagrangian can be written as the form
in (\ref{eq:LA0-1}) again with the CS term given in (\ref{eq:CS term L-1}).

Additionally, we have another term to the ``on-shell'' Lagrangian
which comes from the mass term of $X^{4}$ in this D0-D4 matrix model,

\begin{align}
\frac{\lambda N_{c}M_{KK}}{54\pi}\left(1+\zeta\right)^{3/2}\frac{2}{3}\left(1-\frac{1}{2}\zeta\right)M_{KK}^{2}\mathrm{Tr}\left(X^{4}\right)^{2} & =\frac{\lambda N_{c}M_{KK}^{3}}{3^{4}\pi}\left(1-\frac{1}{2}\zeta\right)\left(1+\zeta\right)^{3/2}\bigg[\frac{1}{4}\left(r_{3}^{4}+\frac{1}{\sqrt{3}}r_{8}^{4}\right)^{2}\nonumber \\
 & \ \ +\frac{1}{4}\left(-r_{3}^{4}+\frac{1}{\sqrt{3}}r_{8}^{4}\right)^{2}+\frac{1}{3}\left(r_{8}^{4}\right)^{2}+\frac{1}{2}\sum_{\rho=1,2,4,5,6,7}\left(r_{\rho}^{4}\right)^{2}\bigg].\label{eq:mass term of X4}
\end{align}
The two- and three-body terms are in the last term of (\ref{eq:mass term of X4})
while the first three terms are related to one baryon potential. So
we need to write the expressions for the off-diagonal $r_{1,2,4,5,6,7}$
to evaluate them.

Basically, the three-body force could be determined in principle by
straightforward calculations from (\ref{eq:kinetic X - 2}) (\ref{eq:kinetic w - 2})
(\ref{eq:mass term of X4}) and (\ref{eq:CS term L-1}). However the
calculations would be very messy and we are less clear about how to
obtain a physical interpretation from the calculations. To clarify
the physical essence, we therefore are going to employ the arguments
as in \cite{key-19} i.e. choose a particular alignment of the baryons.
And the physical essence and significance would be clear by this choice.

\subsection{Hamiltonian for three baryons aligned on a line}

We consider the following condition as in \cite{key-19} for the baryons
aligned on a line,

\begin{equation}
r_{8}^{M}=0,\ \ r_{3}^{M}\equiv r^{M}\neq0,\label{eq:38}
\end{equation}
which means three baryons are located at $x^{M}=r_{3}^{M}/2$, $x^{M}=-r_{3}^{M}/2$
and $x^{M}=0$ respectively. The resultant Lagrangian would be simplified
as

\begin{equation}
L_{A_{0}}=\frac{\lambda N_{c}M_{KK}}{54\pi}\left(1+\zeta\right)^{3/2}\left(L_{1}+L_{2}\right),
\end{equation}
where

\begin{align}
L_{1} & =\frac{162\pi A_{0}^{0}}{\lambda M_{KK}\left(1+\zeta\right)^{3/2}}+\frac{\left(A_{0}^{1}\right)^{2}r^{2}}{2}+\frac{\left(A_{0}^{4}\right)^{2}r^{2}}{8}+\frac{\left(A_{0}^{6}\right)^{2}r^{2}}{8}\nonumber \\
 & \ \ +\left[6\left(A_{0}^{0}\right)^{2}+\left(A_{0}^{1}\right)^{2}+\left(A_{0}^{3}\right)^{2}+\left(A_{0}^{4}\right)^{2}+\left(A_{0}^{6}\right)^{2}+\left(A_{0}^{8}\right)^{2}\right]\rho^{2}\nonumber \\
 & \ \ +\left(A_{0}^{4}A_{0}^{6}+\frac{2A_{0}^{1}A_{0}^{8}}{\sqrt{3}}\right)\rho^{2}u_{0}^{12}+\left(A_{0}^{3}A_{0}^{4}+A_{0}^{1}A_{0}^{6}-\frac{A_{0}^{4}A_{0}^{8}}{\sqrt{3}}\right)\rho^{2}u_{0}^{\left(13\right)}\nonumber \\
 & \ \ +\left(A_{0}^{1}A_{0}^{4}-A_{0}^{3}A_{0}^{6}-\frac{A_{0}^{6}A_{0}^{8}}{\sqrt{3}}\right)\rho^{2}u_{0}^{\left(23\right)}+4A_{0}^{0}\left(A_{0}^{1}u_{0}^{\left(12\right)}+A_{0}^{4}u_{0}^{\left(13\right)}+A_{0}^{6}u_{0}^{\left(23\right)}\right)\rho^{2},\nonumber \\
L_{2} & =\frac{1}{4}\bigg[2\left(A_{0}^{3}\right)^{2}r_{1}^{2}+\frac{1}{2}\left(A_{0}^{4}\right)^{2}r_{1}^{2}+\frac{1}{2}\left(A_{0}^{6}\right)^{2}r_{1}^{2}-A_{0}^{1}A_{0}^{4}r_{1}r_{4}-3A_{0}^{3}A_{0}^{6}r_{1}r_{4}\nonumber \\
 & \ \ -\sqrt{3}A_{0}^{6}A_{0}^{8}r_{1}r_{4}+\frac{1}{2}\left(A_{0}^{1}\right)^{2}r_{4}^{2}+\frac{1}{2}\left(A_{0}^{3}\right)^{2}r_{4}^{2}+\frac{1}{2}\left(A_{0}^{6}\right)^{2}r_{4}^{2}+\sqrt{3}A_{0}^{3}A_{0}^{8}r_{4}^{2}\nonumber \\
 & \ \ +\frac{3}{2}\left(A_{0}^{8}\right)^{2}r_{4}^{2}+3A_{0}^{3}A_{0}^{4}r_{1}r_{6}-A_{0}^{1}A_{0}^{6}r_{1}r_{6}-\sqrt{3}A_{0}^{4}A_{0}^{8}r_{1}r_{6}-A_{0}^{4}A_{0}^{6}r_{4}r_{6}\nonumber \\
 & \ \ +2\sqrt{3}A_{0}^{1}A_{0}^{8}r_{4}r_{6}+\frac{1}{2}\left(A_{0}^{1}\right)^{2}r_{6}^{2}+\frac{1}{2}\left(A_{0}^{3}\right)^{2}r_{6}^{2}+\frac{1}{2}\left(A_{0}^{4}\right)^{2}r_{6}^{2}\nonumber \\
 & \ \ -\sqrt{3}A_{0}^{3}A_{0}^{8}r_{6}^{2}+\frac{3}{2}\left(A_{0}^{8}\right)^{2}r_{6}^{2}\bigg].\label{eq:L1+L2}
\end{align}
For obtaining the expression (\ref{eq:L1+L2}), the terms related
to $r_{3,8}$ and $Y$ have been eliminated since we have used the
following equations

\begin{equation}
r_{3}^{M}r_{1}^{M}=0,\ \left(r_{3}^{M}+\sqrt{3}r_{8}^{M}\right)r_{4}^{M}=0,\ \left(r_{3}^{M}-\sqrt{3}r_{8}^{M}\right)r_{6}^{M}=0.\label{eq:41}
\end{equation}
which could be explicitly shown by the ADHM constraint (\ref{eq:ADHM constraint})
in the expansion of $\left|r\right|\gg\rho$ (See the details of this
expansion in \cite{key-23}). Then we have to evaluate the mass term
of $X^{4}$ in this matrix model, as an explicit result it is

\begin{align}
V_{3-\mathrm{body}}^{\mathrm{mass}} & =\frac{\lambda N_{c}M_{KK}^{3}}{2^{2}3^{4}\pi}\left(1-\frac{1}{2}\zeta\right)\left(1+\zeta\right)^{3/2}\frac{\rho^{6}}{\left|r\right|^{6}}\nonumber \\
 & \ \ \times\bigg\{\mathrm{Tr}\left(rT_{12}\right)\mathrm{Tr}\left[r\left(T_{23}T_{13}-T_{13}T_{23}\right)\right]-2\mathrm{Tr}\left(rT_{31}\right)\mathrm{Tr}\left[r\left(T_{32}T_{12}-T_{12}T_{32}\right)\right]\nonumber \\
 & \ \ -2\mathrm{Tr}\left(rT_{32}\right)\mathrm{Tr}\left[r\left(T_{31}T_{21}-T_{21}T_{31}\right)\right]\bigg\},\label{eq:3-body mass}
\end{align}
where $r=r^{M}\sigma_{M}$ and $T_{ij}=\left(U^{i}\right)^{\dagger}U^{j}-\left(U^{j}\right)^{\dagger}U^{i}=-T_{ji}$.
And we have used (\ref{eq:41}) to simplify (\ref{eq:well-sep ADHM data})
for these aligned baryons to obtain (\ref{eq:3-body mass}) as

\begin{align}
r_{1}^{M}\sigma_{M} & =\frac{1}{\left|r\right|^{2}}\rho^{1}\rho^{2}rT_{21}-\frac{1}{\left|r\right|^{4}}\rho^{1}\rho^{2}\left(\rho^{3}\right)^{2}r\left(T_{32}T_{13}-T_{13}T_{32}\right)+\mathcal{O}\left(1/\left|r\right|^{5}\right),\nonumber \\
r_{4}^{M}\sigma_{M} & =\frac{2}{\left|r\right|^{2}}\rho^{1}\rho^{3}rT_{31}-\frac{1}{\left|r\right|^{4}}\rho^{1}\rho^{3}\left(\rho^{2}\right)^{2}r\left(T_{32}T_{12}-T_{12}T_{32}\right)+\mathcal{O}\left(1/\left|r\right|^{5}\right),\nonumber \\
r_{6}^{M}\sigma_{M} & =\frac{1}{\left|r\right|^{2}}\rho^{2}\rho^{3}rT_{32}-\frac{1}{\left|r\right|^{4}}\rho^{2}\rho^{3}\left(\rho^{1}\right)^{2}r\left(T_{31}T_{21}-T_{21}T_{31}\right)+\mathcal{O}\left(1/\left|r\right|^{5}\right).\label{eq:43}
\end{align}
Note that only the second terms in each right hand side of (\ref{eq:43})
are related to the three-body case while the first terms in the right
hand side of (\ref{eq:43}) equal to the off-diagonal entry of two-body
case. This has been considered in the potential of (\ref{eq:3-body mass})
with taking the classical value as $\rho^{1}=\rho^{2}=\rho^{3}=\rho$
for the leading term in the large $N_{c}$ expansion.

\subsubsection*{C. Integrate out the auxiliary field $A_{0}$}

We also need to solve the equations of motion for $A_{0}$ derived
from Lagrangian (\ref{eq:L1+L2}). By plugging the solution back into
(\ref{eq:L1+L2}), we obtain 

\begin{equation}
L_{A_{0}}=-V,\ \ \ \ V=\sum_{A=1,2,3}V_{1-\mathrm{body}}^{\left(A\right)}+\frac{1}{2}\sum_{A\neq B}V_{2-\mathrm{body}}^{\left(A,B\right)}+V_{3-\mathrm{body}}.\label{eq:44}
\end{equation}

\noindent As (\ref{eq:1,2 body force}) or in \cite{key-08}, the
expressions for one- and two-body potential are

\begin{equation}
V_{1-\mathrm{body}}^{\left(A\right)}=\frac{27\pi N_{c}}{4\lambda M_{KK}\left(1+\zeta\right)^{3/2}}\frac{1}{\rho^{2}},\ \ V_{2-\mathrm{body}}^{\left(A,B\right)}=\frac{27\pi N_{c}}{\lambda M_{KK}\left(1+\zeta\right)^{3/2}}\frac{\left(u_{0}^{\left(AB\right)}\right)^{2}}{\left|r^{\left(AB\right)}\right|^{2}+2\rho^{2}-2\left(u_{0}^{\left(AB\right)}\right)^{2}\rho^{2}}.\label{eq:45}
\end{equation}

\noindent By the condition (\ref{eq:38}) for the aligned baryons

\begin{equation}
\left|r^{\left(12\right)}\right|=r,\ \ \left|r^{\left(13\right)}\right|=\left|r^{\left(23\right)}\right|=r/2,
\end{equation}
and then we will compute the three-body potential $V_{3-\mathrm{body}}$
in (\ref{eq:44}).

As we are going to use the same trick as in \cite{key-19} to solve
the equations of motion for $A_{0}$, we first rewrite the Lagrangian
$L_{1}$ as

\begin{equation}
L_{1}=A^{T}MA+B^{T}A\label{eq:L1 matrix form}
\end{equation}
where

\begin{equation}
A^{T}=\left(A_{0}^{0},A_{0}^{1},A_{0}^{3},A_{0}^{4},A_{0}^{6},A_{0}^{8}\right),\ \ \ B^{T}=\frac{162\pi}{\lambda M_{KK}\left(1+\zeta\right)^{3/2}}\left(1,0,0,0,0,0\right),
\end{equation}
and

\begin{equation}
M=\rho^{2}\left(\begin{array}{cccccc}
6 & 2u_{0}^{\left(12\right)} & 0 & 2u_{0}^{\left(13\right)} & 2u_{0}^{\left(23\right)} & 0\\
2u_{0}^{\left(12\right)} & 1+r^{2}/2\rho^{2} & 0 & u_{0}^{\left(23\right)}/2 & u_{0}^{\left(13\right)}/2 & u_{0}^{\left(12\right)}/\sqrt{3}\\
0 & 0 & 1 & u_{0}^{\left(13\right)}/2 & -u_{0}^{\left(23\right)}/2 & 0\\
2u_{0}^{\left(13\right)} & u_{0}^{\left(23\right)}/2 & u_{0}^{\left(13\right)}/2 & 1+r^{2}/8\rho^{2} & u_{0}^{\left(12\right)}/2 & -u_{0}^{\left(13\right)}/2\sqrt{3}\\
2u_{0}^{\left(23\right)} & u_{0}^{\left(13\right)}/2 & -u_{0}^{\left(23\right)}/2 & u_{0}^{\left(12\right)}/2 & 1+r^{2}/8\rho^{2} & -u_{0}^{\left(23\right)}/2\sqrt{3}\\
0 & u_{0}^{\left(12\right)}/\sqrt{3} & 0 & -u_{0}^{\left(13\right)}/2\sqrt{3} & -u_{0}^{\left(23\right)}/2\sqrt{3} & 1
\end{array}\right).\label{eq:Matrix M}
\end{equation}

There should be another Lagrangian $L_{2}$ for the computation, however
it turns out that Lagrangian $L_{2}$ is not necessary in the next
computation since our computation is in a ``long-distance'' expansion
$\rho\ll r$ and Lagrangian $L_{2}$ is at higher order in this expansion\footnote{As in \cite{key-19}, we have also checked this to confirm that Lagrangian
$L_{2}$ is indeed at higher order for the next computation. However
the computation is lengthy and not necessary for this manuscript thus
it is not presented here.}. 

By (\ref{eq:L1 matrix form}), the solution for the equation of motion
for $A_{0}$ is 

\begin{equation}
A=-\frac{1}{2}M^{-1}B,
\end{equation}
and the Hamiltonian is therefore

\begin{align}
V & =\frac{\lambda M_{KK}N_{c}}{54\pi}\left(1+\zeta\right)^{3/2}\frac{1}{4}B^{T}M^{-1}B\nonumber \\
 & =\frac{3^{5}\pi N_{c}}{2\lambda M_{KK}\left(1+\zeta\right)^{3/2}}\left[M^{-1}\right]_{\left(1,1\right)}.
\end{align}
We obtain the following leading term by expanding in power series
of $\rho^{2}/r^{2}$

\begin{equation}
V=\frac{3^{5}\pi N_{c}}{2\lambda M_{KK}\left(1+\zeta\right)^{3/2}}\left[\frac{1}{6\rho^{2}}+\frac{2\left(u^{\left(1,2\right)}\right)^{2}+8\left(u^{\left(1,3\right)}\right)^{2}+8\left(u^{\left(2,2\right)}\right)^{2}}{9r^{2}}+\frac{4\rho^{2}f_{SI}}{9r^{4}}\right]+\mathcal{O}\left(\rho^{4}/r^{6}\right),\label{eq:52}
\end{equation}
where the function $f_{SI}$ is the spin/isospin phase defined as

\begin{align}
f_{SI} & =\left(u_{0}^{\left(1,2\right)}\right)^{4}-\left(u_{0}^{\left(1,2\right)}\right)^{2}+16\left(u_{0}^{\left(1,3\right)}\right)^{4}-16\left(u_{0}^{\left(1,3\right)}\right)^{2}+16\left(u_{0}^{\left(1,3\right)}\right)^{4}+16\left(u_{0}^{\left(2,3\right)}\right)^{4}-16\left(u_{0}^{\left(2,3\right)}\right)^{2}\nonumber \\
 & +4\left(u_{0}^{\left(1,2\right)}\right)^{2}\left(u_{0}^{\left(1,3\right)}\right)^{2}+4\left(u_{0}^{\left(1,2\right)}\right)^{2}\left(u_{0}^{\left(2,2\right)}\right)^{2}+16\left(u_{0}^{\left(1,3\right)}\right)^{2}\left(u_{0}^{\left(2,3\right)}\right)^{2}-24u_{0}^{\left(1,2\right)}u_{0}^{\left(2,3\right)}u_{0}^{\left(1,3\right)}.
\end{align}
Subtracting the one- and two-body potential (\ref{eq:45}) from (\ref{eq:52}),
we obtain the three-body potential in the expansion of $\rho^{2}/r^{2}$
as

\begin{align}
V_{3-\mathrm{body}}^{A_{0}} & =\frac{216\pi N_{c}\rho^{2}}{\lambda M_{KK}\left(1+\zeta\right)^{3/2}\left|r\right|^{4}}\bigg[\left(u_{0}^{\left(1,2\right)}\right)^{2}\left(u_{0}^{\left(1,3\right)}\right)^{2}+\left(u_{0}^{\left(1,2\right)}\right)^{2}\left(u_{0}^{\left(2,3\right)}\right)^{2}+4\left(u_{0}^{\left(1,3\right)}\right)^{2}\left(u_{0}^{\left(2,3\right)}\right)^{2}\nonumber \\
 & -6u_{0}^{\left(1,2\right)}u_{0}^{\left(2,3\right)}u_{0}^{\left(1,3\right)}\bigg]+\mathcal{O}\left(\rho^{4}/r^{6}\right).\label{eq:V a0}
\end{align}
With the mass term for $X^{4}$ (\ref{eq:3-body mass}), we have the
total three-body potential which is

\begin{equation}
V_{3-\mathrm{body}}=V_{3-\mathrm{body}}^{A_{0}}+V_{3-\mathrm{body}}^{\mathrm{mass}}.\label{eq:55}
\end{equation}
We can evaluate the potential with a three-body baryonic state with
(\ref{eq:V a0}) and (\ref{eq:55}) for any spin or isospin. So we
will choose two different baryonic states as in \cite{key-19} to
study the three-body nuclear potential.

Furthermore, we also have some comments about (\ref{eq:V a0}). As
mentioned that $\rho$ is of order $\mathcal{O}\left(1/\sqrt{\lambda}\right)$,
thus the three-body Hamiltonian is of order $\mathcal{O}\left(1/\lambda^{2}r^{4}\right)$
which is therefore suppressed by $1/\lambda^{2}$. It is also consistent
with \cite{key-20} in which the generic $k$-body potential is of
order $\mathcal{O}\left(1/\lambda^{k-1}r^{2k-2}\right)$ with $k=3$
and $M_{KK}=1$ if setting $\zeta=0$, i.e. no smeared D0-branes.
Additionally, if all the matrices $U^{\left(i\right)}$'s in (\ref{eq:55})
equal to each other which means the ADHM data returns to the 't Hooft
instantons, we would have $u_{0}^{\left(i,j\right)}=1$ and $A_{ij}=0$,
yielding the vanishing three-body potential as same as in (\ref{eq:25})
in this D0-D4/D8 system. Thus obviously it is a consistent check for
the results in the previous section.

\subsubsection*{D. Evaluate the potential with baryonic states}

In this subsection, we are going to compute the spin/isospin dependence
of the three-body short-distance force with our three-body potential
from the D0-D4/D8 matrix model. As a parallel study, we would like
to choose the following two states as in \cite{key-19} which are
\begin{enumerate}
\item three-neutrons with averaged spins.
\item proton-proton-neutron (or proton-neutron-neutron).
\end{enumerate}
The first state is relevant to the dense states of many neutrons as
core of neutron stars or supernovae while the second state is for
the spectrum of Helium-3 nucleus. In some high-density system, the
non-trivial QCD vacuum may affect nuclear force among baryons, as
a description, we would like to use our D0-D4/D8 matrix model to study
the nuclear force with non-trivial QCD vacuum since the number density
of D0-branes in this D0-D4/D8 system is, for example, relevant to
the glueball condensation or CME \cite{key-09,key-10,key-11}.
\begin{description}
\item [{(1)}] three-neutrons with averaged spins.
\end{description}
The single-baryon wave function has been given in (\ref{eq:wave function for nucleon})\footnote{The wave function may be deformed if we consider the baryons or nucleon
with non-zero QCD vacuum. Thus here we use (\ref{eq:wave function for nucleon})
as an ansatz to study the nuclear force with non-zero QCD vacuum.} for protons and neutrons. Since we need neutron states with averaged
spins, thus for any given operators, the appropriate expectation is

\begin{equation}
\left\langle V\right\rangle =\frac{1}{2}\left[\left\langle n\uparrow\left|\mathcal{O}\right|n\uparrow\right\rangle +\left\langle n\downarrow\left|\mathcal{O}\right|n\downarrow\right\rangle \right].\label{eq:56}
\end{equation}
We need to take the expectation value for three baryons for $\mathcal{O}$
being the three-body Hamiltonian. Here we will not anti-symmetrize
the wave function although the nucleons are fermions. We consider
a single baryon case as (\ref{eq:56}), it yields

\begin{equation}
\left\langle V\right\rangle =\int d\Omega_{3}\frac{1}{2}\left[\mathcal{O}\left|\left\langle \vec{a}|n\uparrow\right\rangle \right|^{2}+\mathcal{O}\left|\left\langle \vec{a}|n\downarrow\right\rangle \right|^{2}\right].
\end{equation}
The $d\Omega_{3}$ is the integration over $S^{3}$ by the unit vector
$\vec{a}$. By the wave function (\ref{eq:wave function for nucleon}),
we have

\begin{equation}
\left|\left\langle \vec{a}|n\uparrow\right\rangle \right|^{2}+\left|\left\langle \vec{a}|n\downarrow\right\rangle \right|^{2}=\frac{1}{\pi^{2}}\left[\left(a_{1}\right)^{2}+\left(a_{2}\right)^{2}+\left(a_{3}\right)^{2}+\left(a_{4}\right)^{2}\right]=\frac{1}{\pi^{2}},
\end{equation}
therefore,

\begin{equation}
\left\langle V\right\rangle =\frac{1}{2\pi^{2}}\int d\Omega_{3}\mathcal{O}.\label{eq:59}
\end{equation}
So according to (\ref{eq:59}), with the spin-averaged wave function,
the three-body potential would be,

\begin{align}
\left\langle V_{3-\mathrm{body}}^{A_{0}}\right\rangle _{\mathrm{nnn}\left(\mathrm{spin-averaged}\right)} & =\frac{216\pi N_{c}\rho^{2}}{\lambda M_{KK}\left(1+\zeta\right)^{3/2}\left|r\right|^{4}}\times\nonumber \\
 & \ \ \ \frac{1}{\left(2\pi^{2}\right)^{3}}\int d\Omega_{3}^{\left(1\right)}d\Omega_{3}^{\left(2\right)}d\Omega_{3}^{\left(3\right)}\bigg[\left(u_{0}^{\left(1,2\right)}\right)^{2}\left(u_{0}^{\left(1,3\right)}\right)^{2}+\left(u_{0}^{\left(1,2\right)}\right)^{2}\left(u_{0}^{\left(2,3\right)}\right)^{2}\nonumber \\
 & \ \ \ 4\left(u_{0}^{\left(1,3\right)}\right)^{2}\left(u_{0}^{\left(2,3\right)}\right)^{2}-6u_{0}^{\left(1,2\right)}u_{0}^{\left(2,3\right)}u_{0}^{\left(1,3\right)}\bigg].
\end{align}
And the next computation is quite similar as done in \cite{key-19}.
For example, using $u_{0}=\frac{1}{2}\left(\mathrm{Tr}\left[\left(U^{1}\right)^{\dagger}U^{2}\right]\right)$
and $\left(a_{1}\right)^{2}+\left(a_{2}\right)^{2}+\left(a_{3}\right)^{2}+\left(a_{4}\right)^{2}=1$,
thus for $\left(u_{0}^{\left(1,2\right)}\right)^{2}$ we have

\begin{equation}
u_{0}^{\left(i,j\right)}=\frac{1}{2}\mathrm{Tr}\left[U^{\left(i\right)\dagger}U^{\left(j\right)}\right]=\vec{a}^{\left(i\right)}\cdot\vec{a}^{\left(j\right)},
\end{equation}
where $\vec{a}^{\left(i\right)}$ is unit 4-component vector, pointing
one phase for spin or isospin on $S^{3}$ by the definition of $U$.
Therefore, we can obtain

\begin{align}
\int d\Omega_{3}^{\left(1\right)}\left(u_{0}^{\left(1,2\right)}\right)^{2}=\int d\Omega_{3}^{\left(1\right)}\cos^{2}\theta=\int\cos^{2}\theta\sin^{2}\theta\sin\bar{\theta}d\theta d\bar{\theta}d\phi & =\frac{\pi^{2}}{2},\nonumber \\
\int d\Omega_{3}^{\left(1\right)}d\Omega_{3}^{\left(2\right)}d\Omega_{3}^{\left(3\right)}u_{0}^{\left(1,2\right)}u_{0}^{\left(2,3\right)}u_{0}^{\left(1,3\right)} & =\frac{\pi^{6}}{2},
\end{align}
where $\theta$ is the angle between $\vec{a}^{\left(i\right)}$ and
$\vec{a}^{\left(j\right)}$. Finally we obtain

\begin{equation}
\left\langle V_{3-\mathrm{body}}^{A_{0}}\right\rangle _{\mathrm{nnn}\left(\mathrm{spin-averaged}\right)}=0.
\end{equation}
Therefore we also obtain a vanished three-body potential from the
$A_{0}$ terms with the spin-averaged neutron wave function. 

Similarly, we can obtain the expression of the expectation for $V_{3-\mathrm{body}}^{\mathrm{mass}}$
as

\begin{align}
\left\langle V_{3-\mathrm{body}}^{\mathrm{mass}}\right\rangle _{\mathrm{nnn}\left(\mathrm{spin-averaged}\right)} & =\frac{\lambda N_{c}M_{KK}^{3}}{2^{2}3^{4}\pi}\left(1-\frac{1}{2}\zeta\right)\left(1+\zeta\right)^{3/2}\frac{\rho^{6}}{\left|r\right|^{6}}\nonumber \\
 & \ \ \ \times\frac{1}{\left(2\pi^{2}\right)^{3}}\int d\Omega_{3}^{\left(1\right)}d\Omega_{3}^{\left(2\right)}d\Omega_{3}^{\left(3\right)}\bigg\{\mathrm{Tr}\left(rT_{12}\right)\mathrm{Tr}\left[r\left(T_{23}T_{13}-T_{13}T_{23}\right)\right]\nonumber \\
 & \ \ \ -2\mathrm{Tr}\left(rT_{31}\right)\mathrm{Tr}\left[r\left(T_{32}T_{12}-T_{12}T_{32}\right)\right]-2\mathrm{Tr}\left(rT_{32}\right)\mathrm{Tr}\left[r\left(T_{31}T_{21}-T_{21}T_{31}\right)\right]\bigg\}\nonumber \\
 & =-\frac{\lambda N_{c}M_{KK}^{3}}{2^{2}3^{3}\pi}\left(1-\frac{1}{2}\zeta\right)\left(1+\zeta\right)^{3/2}\frac{\rho^{6}}{\left|r\right|^{6}}\frac{1}{\left(2\pi^{2}\right)^{3}}\nonumber \\
 & \ \ \ \times\int d\Omega_{3}^{\left(1\right)}d\Omega_{3}^{\left(2\right)}d\Omega_{3}^{\left(3\right)}\mathrm{Tr}\left(rT_{21}\right)\mathrm{Tr}\left[r\left(T_{23}T_{13}-T_{13}T_{23}\right)\right].\label{eq:64}
\end{align}
The integration of (\ref{eq:64}) could be performed by using the
polar coordinates and we have used the symmetry for exchanging of
the integration variables. As a result, we have the following integration

\begin{equation}
\frac{1}{\left(2\pi^{2}\right)^{3}}\int d\Omega_{3}^{\left(1\right)}d\Omega_{3}^{\left(2\right)}d\Omega_{3}^{\left(3\right)}\mathrm{Tr}\left(rT_{21}\right)\mathrm{Tr}\left[r\left(T_{23}T_{13}-T_{13}T_{23}\right)\right]=-8\left|\vec{r}\right|^{2}.
\end{equation}
Therefore we obtain the expectation from (\ref{eq:64}) which is

\begin{equation}
\left\langle V_{3-\mathrm{body}}^{\mathrm{mass}}\right\rangle _{\mathrm{nnn}\left(\mathrm{spin-averaged}\right)}=\frac{2^{-1/2}3^{15/2}\pi^{2}N_{c}}{\lambda^{2}M_{KK}^{3}\left(1-\frac{1}{2}\zeta\right)^{1/2}\left(1+\zeta\right)^{3}\left|\vec{r}\right|^{4}},\label{eq:66}
\end{equation}
where the three-dimensional vector $\vec{r}$ specifies the inter-baryon
distance in our space. The four-dimensional distance could be identified
as three-dimensional distance since we can choose the classical value
for the four-dimensional component $r^{4}$ of $r^{M}=\left(\vec{r},r^{4}\right)$
vanished at leading order in $1/N$ expansion. And in (\ref{eq:66}),
we have substituted the classical value of $\rho$, which is $\rho=2^{-1/4}3^{7/4}\sqrt{\pi}\lambda^{-1/3}M_{KK}^{-1}\left(1-\frac{1}{2}\zeta\right)^{-1/4}\left(1+\zeta\right)^{-3/4}$
in \cite{key-08}, for two-flavor case also for the leading order
in the $1/N_{c}$ expansion. So we obtain the total three-body potential
in our D0-D4/D8 system as,

\begin{align}
\left\langle V_{3-\mathrm{body}}\right\rangle _{\mathrm{nnn}\left(\mathrm{spin-averaged}\right)} & =\left\langle V_{3-\mathrm{body}}^{A_{0}}\right\rangle _{\mathrm{nnn}\left(\mathrm{spin-averaged}\right)}+\left\langle V_{3-\mathrm{body}}^{\mathrm{mass}}\right\rangle _{\mathrm{nnn}\left(\mathrm{spin-averaged}\right)}\nonumber \\
 & =\frac{2^{-1/2}3^{15/2}\pi^{2}N_{c}}{\lambda^{2}M_{KK}^{3}\left(1-\frac{1}{2}\zeta\right)^{1/2}\left(1+\zeta\right)^{3}\left|\vec{r}\right|^{4}}.\label{eq:67}
\end{align}
This three-body force is obtained by considering effect of non-trivial
QCD vacuum in the D0-D4/D8 system and also with averaged spin. The
three-body force is suppressed if compared to the two-body potential
(\ref{eq:Nuclear potential zero order}) for large $\lambda$. As
in \cite{key-20}, our three-body potential is also a generic hierarchy
between $N+1$- to $N$-body potential. $M_{KK}$ is the energy scale
for the dual QCD-like field theory, and our calculations are as well
valid at short distances. However if we focus on the factor $\left(1-\frac{1}{2}\zeta\right)^{1/2}$,
it implies that the three-body potential is totally complex with $\zeta>2$,
which is nothing but our constraint for stable baryonic state in this
system. We will discuss about it in details in the final section.
\begin{description}
\item [{(2)}] proton-proton-neutron (or proton-neutron-proton).
\end{description}
In this subsection, let us evaluate the three-body potential with
the state of proton-proton-neutron (which is also a same calculation
for the case of proton-neutron-proton). We will consider the three-nucleon
state with a total spin $1/2$ and a total isospin $1/2$. We can
use the following state to represent a proton-proton-neutron state
with the third component of the total isospin $+1/2$, 

\begin{align}
\frac{1}{\sqrt{6}}\bigg[ & |p\uparrow>_{1}|p\downarrow>_{2}|n\downarrow>_{3}-|p\downarrow>_{1}|p\uparrow>_{2}|n\uparrow>_{3}-|p\uparrow>_{1}n\uparrow>_{2}|p\downarrow>_{3}\nonumber \\
+ & |p\downarrow>_{1}|n\uparrow>_{2}|p\uparrow>_{3}-|n\uparrow>_{1}|p\downarrow>_{2}|p\uparrow>_{3}+|n\uparrow>_{1}|p\uparrow>_{2}|p\downarrow>_{3}\bigg].
\end{align}
The next calculations are straightforward and similar to what we have
done for three-neutrons with averaged spins. With the following integrals

\begin{align}
\int d\Omega_{3}^{\left(1\right)}d\Omega_{3}^{\left(2\right)}d\Omega_{3}^{\left(3\right)}\left|\psi\left(\vec{a}_{1},\vec{a}_{2},\vec{a}_{3}\right)\right|^{2}\left(u_{0}^{\left(1,2\right)}\right)^{2}\left(u_{0}^{\left(1,3\right)}\right)^{2} & =\frac{1}{36},\nonumber \\
\int d\Omega_{3}^{\left(1\right)}d\Omega_{3}^{\left(2\right)}d\Omega_{3}^{\left(3\right)}\left|\psi\left(\vec{a}_{1},\vec{a}_{2},\vec{a}_{3}\right)\right|^{2}u_{0}^{\left(1,2\right)}u_{0}^{\left(2,3\right)}u_{0}^{\left(1,3\right)} & =\frac{1}{36},\nonumber \\
\int d\Omega_{3}^{\left(1\right)}d\Omega_{3}^{\left(2\right)}d\Omega_{3}^{\left(3\right)}\left|\psi\left(\vec{a}_{1},\vec{a}_{2},\vec{a}_{3}\right)\right|^{2}\mathrm{Tr}\left(rT_{21}\right)\mathrm{Tr}\left[r\left(T_{23}T_{13}-T_{13}T_{23}\right)\right] & =-\frac{320}{27}\left|\vec{r}\right|^{2}.
\end{align}
With these formulas, we have

\begin{align}
\left\langle V_{3-\mathrm{body}}^{A_{0}}\right\rangle _{\mathrm{ppn}} & =0,\nonumber \\
\left\langle V_{3-\mathrm{body}}^{\mathrm{mass}}\right\rangle _{\mathrm{ppn}} & =\frac{2^{5/2}3^{9/2}5\pi^{2}N_{c}}{\lambda^{2}M_{KK}^{3}\left(1-\frac{1}{2}\zeta\right)^{1/2}\left(1+\zeta\right)^{3}\left|\vec{r}\right|^{4}}.
\end{align}
Therefore we have the total three-body potential for proton-proton-neutron
which is

\begin{align}
\left\langle V_{3-\mathrm{body}}\right\rangle _{\mathrm{ppn}} & =\left\langle V_{3-\mathrm{body}}^{A_{0}}\right\rangle _{\mathrm{ppn}}+\left\langle V_{3-\mathrm{body}}^{\mathrm{mass}}\right\rangle _{\mathrm{ppn}}\nonumber \\
 & =\frac{2^{5/2}3^{9/2}5\pi^{2}N_{c}}{\lambda^{2}M_{KK}^{3}\left(1-\frac{1}{2}\zeta\right)^{1/2}\left(1+\zeta\right)^{3}\left|\vec{r}\right|^{4}}.\label{eq:71}
\end{align}
This three-body potential is positive which means there is a repulsive
three-body force at short distances. The computation for three other
wave functions $\left(+1/2,-1/2\right)$, $\left(-1/2,+1/2\right)$
and $\left(-1/2,-1/2\right)$ of the third component of the spin and
isospin, is the same as the current computation for $\left(+1/2,+1/2\right)$.
And the result turns out to be the same as (\ref{eq:71}) since our
matrix model (\ref{eq:Matrix action}) is $SO\left(3\right)$ invariance
for rotational symmetry and $SU\left(2\right)$ invariance for isospin
symmetry.

\section{Summary and discussion}

We proposed a matrix model with $U\left(k\right)$ gauge symmetry
in \cite{key-08} for $k$-body baryon systems with non-trivial QCD
vacuum. And in this paper, we use this matrix model to compute the
three-body force for baryons at short distances. We find the result
includes some effects, maybe such as in glueball condensation and
CME, from the non-trivial QCD vacuum. We derived the matrix model
by using the standard technique from gauge/string duality (also the
AdS/CFT correspondence), thus our matrix model is not a phenomenological
model. Precisely, the matrix model is a low-energy effective theory
for the baryon vertex, which is denoted as a D4'-brane, in the D0-D4/D8
holographic system of large $N_{c}$ QCD with non-trivial vacuum.
Consequently, we can compute the $k$-body baryon potentials for arbitrary
number of $k$ with this framework by considering the non-trivial
QCD vacuum.

Our computation is parallel to \cite{key-19} thus is straightforward.
We took $k=3$ for the case of three baryons, i.e. the $U\left(3\right)$
matrix model and evaluate the Hamiltonian with a quantum three-body
state which is a tensor product of single-baryon states. Then the
potential intrinsic to the three-body case is obtained after subtracting
the one- and two-body contributions. However our calculations are
valid only at short distances, i.e. $1/\sqrt{\lambda}M_{KK}\ll\left|r\right|\ll1/M_{KK}$
where $\lambda$ is the 't Hooft coupling constant\footnote{If we fit pion decay constant with $\lambda$, then $M_{KK}$ would
be $\mathcal{O}\left(1\text{ \textemdash\ 0.5}\right)$GeV when it
fit with the mass of baryon or meson \cite{key-16,key-17,key-24}.}. As two typical and explicit examples, we choose (1) three-neutrons
with averaged spins and (2) proton-proton-neutron (or proton-neutron-proton),
and in both cases the baryons or nucleons are aligned on a line with
equal spacings. We obtain the resultant three-body potentials for
baryons in (\ref{eq:67}) and (\ref{eq:71}), both of which are positive
(i.e. repulsive) and modified by the appearance of the smeared D0-branes
(i.e. the non-trivial QCD vacuum). And as a quick check, all our results
would return to \cite{key-19} if setting $\zeta=0$, i.e. no smeared
D0-branes.

Furthermore, we would like to give some more comments to our results
and discuss the importance of them. According to the form of the wave
function (\ref{eq:wave function for nucleon}), our results (\ref{eq:67})
hold also for the case of three-protons since the matrix action is
$SU\left(2\right)$ invariance for isospin. So the results (\ref{eq:67})
hold if all three baryons or nucleons have the same flavor. Therefore,
the three-body potential for proton-neutron-neutron takes the same
form as (\ref{eq:67}), which implies the additional repulsive three-body
force may exist in addition to two-body force and be affected by some
effects from non-trivial QCD vacuum at short distances. Besides, the
three-body potentials obtained in (\ref{eq:67}) and (\ref{eq:71})
are suppressed if compared with the two-body potential (\ref{eq:Nuclear potential zero order}).
At short distances, i.e. $1/\sqrt{\lambda}M_{KK}\ll\left|r\right|\ll1/M_{KK}$,
the suppression factor $1/\lambda\left(rM_{KK}\right)^{2}$ is small
which makes our computation valid.

On the other hand, as a difference from the original Sakai-Sugimoto
model, we find the resultant three-body potentials (\ref{eq:67})
and (\ref{eq:71}) are totally complex if $\zeta>2$. It corresponds
to the constraint for stable baryonic states in the D0-D4/D8 holographic
system discussed in \cite{key-08,key-11} for two-body case. During
our computation, the three-body potentials (\ref{eq:67}) and (\ref{eq:71})
actually come from the mass term in the matrix action (\ref{eq:Matrix action})
since the contributions from other terms vanish. From the matrix action
(\ref{eq:Matrix action}), it is obvious to see that the matrix model
describe an unstable system if $\zeta>2$ (i.e. a quantum mechanical
system with complex mass term), that is the reason that the computation
does not depend on the number density of the smeared D0-branes in
the (\ref{eq:25}) from ``classical treatment'', while our results
(\ref{eq:67}) and (\ref{eq:71}) are also consistent with these.
Therefore according to our results in \cite{key-08} and the three-body
force (\ref{eq:67}) (\ref{eq:71}), it implies that the constraint
for stable baryonic state may hold in the methods for $N$-body case.
Besides, if comparing our results (\ref{eq:67}) and (\ref{eq:71})
with the two-body force (\ref{eq:Nuclear potential zero order}) and
(\ref{eq:Nuclear potential leading order}), we find the three-body
force would be going to infinity as $\zeta\rightarrow1/2$. This implies
the three-body force would become dominant if the non-trivial QCD
vacuum is too important to be neglected, which is also different from
\cite{key-19}.

With $\zeta<2$, these three-body forces would also become stronger
if the distances get shorter. As a result, three-body forces give
additional repulsive potential at short distances if the neutrons
are highly dense. As mentioned, the effects from non-trivial QCD vacuum,
for example in the glueball condensation or CME, may also play the
important roles in such high-density matter. Physically, the mass
spectrum of mesons is modified by considering the effects from non-trivial
QCD vacuum with the D0-D4/D8 holographic system as discussed in \cite{key-09},
so the potential of the interaction among baryons would also be modified
since the nucleons interact with each other by exchanging such mesons,
which is also consistent with \cite{key-08,key-11}. In the viewpoint
of dual field theory, adding smeared D0-branes equals to add non-zero
$\theta\mathrm{Tr}\left(F_{\mu\nu}\tilde{F}^{\mu\nu}\right)$ term
to the action. With this term, the propagator derived in the dual
quantum field theory is modified, thus yielding the modified three-body
potential. Additionally, we also find the three-body forces for proton-proton-neutron
and proton-neutron-neutron are all positive i.e. repulsive, thus our
result seems also responsible for Helium and Triton if the effects
from the non-trivial vacuum are considered.

However, our results are as examples limited to three baryons on a
line and only valid at short distances since the calculations for
three baryons with generic positions are too messy to get the physical
significance. So our result is not conclusive enough for those interests
listed above but suggestive. Therefore, there is still a long way
from holographic model with an underlying theory towards real-world
nuclear matters.

\section*{Acknowledgments}

This work is inspired by a seminar given by Chao Wu on his works \cite{key-09,key-11}
and also as an extension to our previous works \cite{key-08,key-25,key-26}
in USTC. And we would like to thank Prof. Qun Wang and Dr. Chao Wu
for helpful discussions.


\begin{thebibliography}{10}
\bibitem[1]{key-01}H. Leutwyler, Phys. Lett. B 96 (1980) 154; Nucl.
Phys. B 179 (1981) 129..

\bibitem[2]{key-02}P. Minkowski, Nucl. Phys. B 177 (1981) 203..

\bibitem[3]{key-03}C. A. Flory, Phys. Rev. D 28 (1983) 1425.

\bibitem[4]{key-04}P. van Baal, Commun. Math. Phys. 94 (1984) 397

\bibitem[5]{key-05}G. V. Efimov, A. C. Kalloniatis, S. N. Nedelko,
Phys. Rev. D 59 (1999) 014026 {[}hep-th/9806165{]}.

\bibitem[6]{key-06}J. Liao, ``Chiral Magnetic Effect in Heavy Ion
Collisions'', {[}arXiv:1601.00381{]}.

\bibitem[7]{key-07}D. E. Kharzeev, J. Liao, S. A. Voloshin, G. Wang,
``Chiral Magnetic Effect in High-Energy Nuclear Collisions --- A
Status Report'', {[}arXiv:1511.04050{]}.

\bibitem[8]{key-08}S. Li, T. Jia, ``Matrix model and Holographic
Baryons in the D0-D4 background'', Phys. Rev. D 92 (2015) 046007,
{[}arXiv:1506.00068{]}.

\bibitem[9]{key-09}C. Wu, Z. Xiao, D. Zhou, ``Sakai-Sugimoto model
in D0-D4 background'', Phys.Rev.D.88 (2013) 026016.

\bibitem[10]{key-10}K. Suzuki, \textquotedblleft D0-D4 system and
QCD\_\{3+1\}'', Phys.Rev. D63 (2001) 084011, {[}arXiv:hep-th/0001057{]}.

\bibitem[11]{key-11}W. Cai, C. Wu, Z. Xiao, ``Baryons in the Sakai-Sugimoto
model in the D0-D4 background'', Phys.Rev. D90 (2014) 106001, {[}arXiv:1410.5549{]}.

\bibitem[12]{key-12}K. Hashimoto, N. Iizuka, P. Yi, \textquotedblleft A
Matrix Model for Baryons and Nuclear Forces\textquotedblright{} ,
JHEP 10 (2010) 003 {[}arXiv:1003.4988{]} .

\bibitem[13]{key-13}J. M. Maldacena, \textquotedblleft The large
N limit of superconformal field theories and supergravity\textquotedblright ,
Adv. Theor. Math. Phys. 2, 231 (1998) {[}Int. J. Theor. Phys. 38,
1113 (1999){]} {[}arXiv:hep-th/9711200{]}.

\bibitem[14]{key-14}S. S. Gubser, I. R. Klebanov and A. M. Polyakov,
\textquotedblleft Gauge theory correlators from non-critical string
theory\textquotedblright , Phys. Lett. B 428, 105 (1998) {[}arXiv:hep-th/9802109{]}.

\bibitem[15]{key-15}E. Witten, \textquotedblleft Anti-de Sitter space
and holography\textquotedblright , Adv. Theor. Math. Phys. 2, 253
(1998) {[}arXiv:hep-th/9802150{]}.

\bibitem[16]{key-16}T. Sakai, S. Sugimoto, \textquotedblleft Low
energy hadron physics in holographic QCD\textquotedblright , Prog.
Theor. Phys. 113, 843 (2005) {[}arXiv:hep-th/0412141{]}.

\bibitem[17]{key-17}T. Sakai, S. Sugimoto, \textquotedblleft More
on a holographic dual of QCD\textquotedblright , Prog. Theor. Phys.
114, 1083 (2005) {[}arXiv:hep-th/0507073{]}.

\bibitem[18]{key-18}E. Witten, \textquotedblleft Baryons and branes
in anti de Sitter space\textquotedblright , JHEP 9807, 006 (1998)
{[}arXiv:hep-th/9805112{]}.

\bibitem[19]{key-19}K. Hashimoto, N. Iizuka, ``Three-Body Nuclear
Forces from a Matrix Model'', JHEP 11 (2010) 058, {[}arXiv:1005.4412{]}.

\bibitem[20]{key-20}K. Hashimoto, N. Iizuka, T. Nakatsukasa, ``N-Body
Nuclear Forces at Short Distances in Holographic QCD'', Phys. Rev.
D 81 (2010) 6003, {[}arXiv:0911.1035{]}.

\bibitem[21]{key-21}M. F. Atiyah, N. J. Hitchin, V. G. Drinfeld,
Yu. I. Manin, \textquotedblleft Construction of instantons\textquotedblright ,
Phys. Lett. A 65, 185 (1978).

\bibitem[22]{key-22}K. Hashimoto, T. Sakai, S. Sugimoto, ``Nuclear
Force from String Theory'' , Prog. Theor. Phys. \textbf{122} (2009)
427-476, {[}arXiv:0901.4449{]}.

\bibitem[23]{key-23}N. H. Christ, E. J. Weinberg, N. K. Stanton,
\textquotedblleft General self-dual Yang-Mills solutions,\textquotedblright{}
Phys. Rev. D 18 (1978) 2013.

\bibitem[24]{key-24}H. Hata, T. Sakai, S. Sugimoto, S. Yamato, \textquotedblleft Baryons
from instantons in holographic QCD\textquotedblright , Prog. Theor.
Phys. 117, 1157 (2007), {[}arXiv:hep-th/0701280{]}.

\bibitem[25]{key-25}S. Li, A. Schmitt, Q. Wang, ``From holography
towards real-world nuclear matter'', Phys.Rev. D92 (2015) 026006,
{[}arXiv:1505.04886{]}.

\bibitem[26]{key-26}S. Li, ``Baryon Transition in Holographic QCD'',
{[}arXiv:1509.06914{]}.\end{thebibliography}
\end{document}